\documentclass{aastex631}

\usepackage{soul}
\usepackage{longfigure}
\usepackage{rotating}
\usepackage{booktabs}
\usepackage[toc,page]{appendix}
\submitjournal{PSJ}

\begin{document}

\title{Size Constraints for the Pre-atmospheric Parent Bodies of Ordinary Chondrites}

\correspondingauthor{Juan A. Sanchez}
\email{jsanchez@psi.edu}

\author{Juan A. Sanchez}
\altaffiliation{Visiting Astronomer at the Infrared Telescope Facility, which is operated by the University of Hawaii under Cooperative Agreement no. NNX-08AE38A with 
the National Aeronautics and Space Administration, Science Mission Directorate, Planetary Astronomy Program.}
\affiliation{Planetary Science Institute, 1700 East Fort Lowell Road, Tucson, AZ 85719, USA}

\author{David C. Cantillo}
\altaffiliation{Visiting Astronomer at the Infrared Telescope Facility, which is operated by the University of Hawaii under Cooperative Agreement no. NNX-08AE38A with 
the National Aeronautics and Space Administration, Science Mission Directorate, Planetary Astronomy Program.}
\affiliation{Lunar and Planetary Laboratory, University of Arizona, 1629 E University Blvd, Tucson, AZ 85721-0092}

\author{Adam Battle}
\altaffiliation{Visiting Astronomer at the Infrared Telescope Facility, which is operated by the University of Hawaii under Cooperative Agreement no. NNX-08AE38A with 
the National Aeronautics and Space Administration, Science Mission Directorate, Planetary Astronomy Program.}
\affiliation{Lunar and Planetary Laboratory, University of Arizona, 1629 E University Blvd, Tucson, AZ 85721-0092}

\author{Miren Miranda}
\altaffiliation{Visiting Astronomer at the Infrared Telescope Facility, which is operated by the University of Hawaii under Cooperative Agreement no. NNX-08AE38A with 
the National Aeronautics and Space Administration, Science Mission Directorate, Planetary Astronomy Program.}
\affiliation{Lunar and Planetary Laboratory, University of Arizona, 1629 E University Blvd, Tucson, AZ 85721-0092}

\author{Benjamin N. L. Sharkey}
\altaffiliation{Visiting Astronomer at the Infrared Telescope Facility, which is operated by the University of Hawaii under Cooperative Agreement no. NNX-08AE38A with 
the National Aeronautics and Space Administration, Science Mission Directorate, Planetary Astronomy Program.}
\affiliation{Department of Astronomy, University of Maryland 4296 Stadium Dr. PSC (Bldg 415) Rm 1113 College Park, MD 20742-2421, USA}

\author{Tanner S. Campbell}
\affiliation{Lunar and Planetary Laboratory, University of Arizona, 1629 E University Blvd, Tucson, AZ 85721-0092}

\author{Rogerio Deienno}
\affiliation{Department of Space Studies, Southwest Research Institute, 1050 Walnut Street, Suite 300 Boulder, CO 80302, USA}

\author{William F. Bottke}
\affiliation{Department of Space Studies, Southwest Research Institute, 1050 Walnut Street, Suite 300 Boulder, CO 80302, USA}

\author{David Nesvorn\'{y}}
\affiliation{Department of Space Studies, Southwest Research Institute, 1050 Walnut Street, Suite 300 Boulder, CO 80302, USA}

\author{Vishnu Reddy}
\altaffiliation{Visiting Astronomer at the Infrared Telescope Facility, which is operated by the University of Hawaii under Cooperative Agreement no. NNX-08AE38A with 
the National Aeronautics and Space Administration, Science Mission Directorate, Planetary Astronomy Program.}
\affiliation{Lunar and Planetary Laboratory, University of Arizona, 1629 E University Blvd, Tucson, AZ 85721-0092}

\begin{abstract}

The study of S-complex near-Earth objects (NEOs), the parent bodies of ordinary chondrites, has shown that they are dominated by asteroids with LL chondrite-like 
compositions. This is surprising because among the three subtypes of ordinary chondrites (H, L, and LL), LL chondrites are the least common, representing only 10\% 
of all ordinary chondrite falls. This discrepancy has been attributed to the size of the NEOs studied, which are likely too large to be the immediate precursors of the 
meteorites that fall on Earth. To test this hypothesis, we obtained near-infrared spectra (0.7-2.5 $\mu$m) of a group of objects with absolute magnitudes 20.0 $\leq$ $H$ $\leq$ 29.2 (sizes $\sim$4-343 m). The sample was divided into subgroups based on their $H$ value, and the composition of the asteroids was determined. We found that the dominance of LL chondrite-like objects disappears at sizes of $\sim$31-49 m. At this size range, asteroids 
with L chondrite-like compositions become dominant, matching the fraction of L chondrite meteorite falls. In contrast, the fraction of H chondrite-like NEOs was found to be much lower than the proportion of H chondrite falls, even among the smallest objects. We determined an upper size limit of 
$\sim$18 m for the parent bodies of these meteorites. The same upper limit was established for the pre-atmospheric parent bodies of LL chondrites. These results constitute the first observational evidence for a size dependence in the composition of S-complex NEOs.

\end{abstract}

\keywords{Asteroids; Near-Earth Objects; Spectroscopy}

\section{Introduction} \label{sec:intro}

Silicate-rich asteroids are the most common type of observed near-Earth objects (NEOs), making up over 60\% of this 
population \citep[e.g.,][]{2019Icar..324...41B, 2024PSJ.....5..131S}.  These asteroids, which constitute the so-called S-complex, have been linked to 
ordinary chondrites based on spectroscopic observations and returned samples \citep[e.g.,][]{1993Icar..106..573G, 2011ScienceNakamura, 2013Icar..222..273D, 2019Icar..324...41B, 2024PSJ.....5..131S}. Ordinary 
chondrites are the most common type of meteorites, representing about 81\% of all meteorite falls (The Meteoritical Bulletin). They are mostly composed of olivine, pyroxene, plagioclase feldspar, and small amounts of metal and 
sulfides \citep{2002aste.book..653B, 2006mess.book...19W}. These meteorites are divided into three subtypes (H, L, LL) based on the abundance of Fe and the ratio of 
metallic Fe (Fe$^{0}$) to oxidized Fe (FeO). Among ordinary chondrite falls, L chondrites are the most common (47\%), followed by H chondrites (43\%), with 
LL chondrites being the least common (10\%). 

The study of S-complex NEOs, with diameters extending from $\sim$100 m to kilometer-sized bodies, has shown that they are dominated by objects with LL chondrite-like compositions, with proportions ranging from $\sim$51 to 66\% \citep[e.g.,][]{2008Natur.454..858V, 2013Icar..222..273D, 2019Icar..324...41B, 2024PSJ.....5..131S}. In this size range, asteroids with H chondrite-like compositions represent $\sim$15-29\%, and those with L chondrite-like composition $\sim$10-20\% of this population. This difference in proportions with respect to ordinary chondrite falls has been attributed to the size of the NEOs studied, which are likely too large to be the immediate (pre-atmospheric) 
precursors of the meteorites that fall on Earth. Recent asteroid breakups in the asteroid belt 
dominate the flux of small objects, and young asteroid families tend to exhibit steep size-frequency distributions down to meter-sized bodies \citep{2024Natur.634..566B, 2024Natur.634..561M}. The Yarkovsky effect is more efficient at transporting these small objects than larger asteroids to resonances that increase their orbital eccentricity. As a result, young asteroid families located closer to major resonances contribute more efficiently to the NEO population.

High-inclination NEOs with H chondrite-like compositions are thought to have originated from the Phocaea, Maria or Nele family \citep{2024Natur.634..566B, 2025M&PS...60..928J}. For NEOs with low inclination and semimajor axis greater than 2.5 au, the Karin and Koronis$_{2}$ families have been identified as 
their possible source \citep{2024Natur.634..566B}, although \cite{2026AJ....171..169V} argued that Koronis$_{2}$ is older than previously estimated, 
which would weaken the link between the family age and the cosmic-ray exposure ages of the H chondrites.

The Juno family, located between the 3:1 and 5:2 resonances, has been proposed as the main source of kilometer-sized L chondrite-like NEOs, whereas 
the Massalia$_{2}$ family, which lies between the $\nu_{6}$ and 3:1 resonances, is considered a viable source for L chondrite meteorites 
\citep{2024Natur.634..566B, 2024Natur.634..561M}. In particular, \cite{2024Natur.634..561M} predicted that the fraction of NEOs with L chondrite-like 
composition must increase at sizes smaller than $\sim$100 m due to the transport timescales from the Massalia family to the $\nu_{6}$ resonance. This 
means that small L chondrite-like NEOs should preferentially originate from this resonance and exhibit lower semimajor axes, while larger objects should be 
associated with the 3:1 resonance. It should be noted that within this family, objects with H chondrite-like compositions have also been found \citep{2020AAS...23527707H}. In addition, other asteroid families have been proposed as potential sources of L chondrites. These include the Gefion family, located near the 5:2 resonance, and the 
low-inclination Hertha family, which lies between the $\nu_{6}$ and 3:1 resonances \citep{2009Icar..200..698N, 2025M&PS...60..928J}. 
Several objects with L chondrite-like compositions have been identified in the Gefion family \citep[e.g.,][]{2025MNRAS.537.3145M}; however, for the Hertha family the compositional data is sparse \citep{2015Icar..252..199D, 2025M&PS...60..928J}.

For LL chondrite-like NEOs, the Flora family in the inner part of the asteroid belt and the S-type component of the Nysa family have been identified as their 
dominant source \citep[e.g.,][]{2008Natur.454..858V, 2024Natur.634..566B, 2026A&A...598A..11M}. The Flora family appears to supply primarily 
equilibrated LL chondrites, while the Nysa family supplies low-petrologic-type material \citep{2026A&A...598A..11M}. 

To date, there is no observational evidence of a size dependence in the composition of S-complex NEOs that could explain the discrepancy between these objects and ordinary chondrite falls. This is mainly due to the difficulties in 
obtaining high-quality near-infrared (NIR) spectra of small NEOs from which their composition can be derived. In this work, we study a group of NEOs spanning a wide range of sizes to determine whether such a size dependence exists. Furthermore, we explore whether the smallest objects in our sample exhibit preferential source regions 
relative to the largest NEOs.

\section{Observations and Data Reduction} \label{sec:obs}

The NEOs  were observed with the SpeX instrument \citep{2003PASP..115..362R} on NASA Infrared Telescope Facility (IRTF) between May-2021 and July-2025. NIR spectra (0.7-2.5 $\mu$m) were obtained in low-resolution 
($R$$\sim$150) prism mode with a 0.8” slit width. During the observations, the slit was oriented along the parallactic angle in order to minimize the effects of differential atmospheric refraction. Spectra were obtained in 
two different slit positions (A-B) following the sequence ABBA.  Integration times varied between 120 and 200 s depending on the visual magnitude of the asteroid and weather conditions. In 
order to correct the telluric bands from the asteroid spectra, a G-type local extinction star was observed before and after the asteroid. NIR spectra of a solar analog were also obtained to correct for possible spectral slope variations that 
could be introduced by the use of a nonsolar local extinction star. For all NEOs, guiding was done using the MIT Optical Rapid Imaging System (MORIS) instrument and a 0.7 $\mu$m cut-on dichroic. Calibration images including flat fields and argon arc lamp 
spectra were also obtained. Observational circumstances for the 25 NEOs are presented in Table 1. 

NIR spectra were reduced using the IDL-based software Spextool \citep{2004PASP..116..362C} and several Python scripts following the same procedure described in \cite{2013Icar..225..131S}. 
The data reduction procedure includes the following steps: (1) sky background removal by subtracting the A-B image pairs, (2) flat-fielding, (3) cosmic-ray and spurious hit removals, (4) wavelength 
calibration, (5) division of asteroid spectra by the spectra of the local extinction star and solar analog star, and (6) co-adding of individual spectra. The spectra of the 25 NEOs are shown in the 
Appendix.

\startlongtable
\begin{deluxetable*}{ccccccccc}

\tablecaption{\label{t:Table1} {\small Observational Circumstances. The columns in this table are: object number and designation, date, absolute magnitude ($H$), 
$V$ magnitude ($V$), phase angle ($\alpha$), heliocentric distance ($r$), airmass and solar analog used.}}

\tablewidth{0pt}

\tablehead{Number&Designation&Date (UT)&$H$ (mag)&$V$ (mag)&$\alpha$ $(^{\circ})$&$r$ (au)&Airmass&Solar Analog \\}

\startdata
462959&2011 DU&12-April-2025&21.1&16.7&16.4&1.08&1.04&SAO 120107 \\
483508&2003 CR1&22-January-2024&20.0&17.2&71.7&1.01&1.15&SAO 120107 \\
811221&2022 FR3&29-March-2025&22.6&16.5&60.1&1.01&1.01&SAO 93936 \\
826663&2021 PT&23-August-2021&21.9&17.6&76.8&1.02&1.32&SAO 120107 \\
826936&1998 HH49&17-October-2023&21.4&13.1&58.0&1.0&1.52&SAO 109542 \\
&2013 WV44&22-June-2023&22.9&18.3&52.0&1.05&1.1&SAO 120107 \\
&2019 UT6&15-November-2023&21.9&17.4&21.0&1.06&1.16&SAO 109542 \\
&2020 UQ3&16-July-2023&24.1&17.2&66.0&1.02&1.07&SAO 120107 \\
&2021 JG1&20-May-2021&24.8&18.4&18.0&1.04&1.22&SAO 120107 \\
&2021 SZ4&20-October-2023&20.5&16.9&23.7&1.09&1.04&SAO 109542 \\
&2023 PM&20-August-2023&24.0&18.1&57.0&1.03&1.02&SAO 83469 \\
&2023 QC&20-August-2023&24.7&15.9&9.6&1.03&1.32&SAO 83469 \\
&2023 UH5&25-October-2023&25.2&17.3&11.0&1.01&1.44&SAO 93936 \\
&2023 UT&14-November-2023&21.8&17.9&25.8&1.07&1.2&SAO 93936 \\
&2023 VD6&16-December-2023&21.6&17.2&34.2&1.04&1.2&SAO 93936 \\
&2023 VR4&15-November-2023&23.0&17.5&23.6&1.03&1.25&SAO 109542 \\
&2024 BH&09-February-2024&23.2&18.0&30.1&1.03&1.15&SAO 93936 \\
&2024 MT1&07-July-2024&23.3&15.4&45.0&1.03&1.2&SAO 83469 \\
&2024 OM1&05-August-2024&24.2&16.0&24.5&1.03&1.53&SAO 83469 \\
&2024 QE2&03-September-2024&27.5&17.8&30.0&1.01&1.03&SAO 93936 \\
&2024 WB&02-December-2024&21.1&16.3&7.1&1.07&1.03&SAO 93936 \\
&2025 FU5&27-May-2025&23.2&16.9&15.8&1.05&1.26&SAO 120107 \\
&2025 HM&28-April-2025&22.6&17.6&37.9&1.04&1.25&SAO 120107 \\
&2025 OL1&27-July-2025&25.0&17.7&24.3&1.03&1.07&SAO 120107 \\
&2025 OW&27-July-2025&23.5&16.7&49.0&1.03&1.07&SAO 120107 \\
\enddata
\end{deluxetable*}

\section{Sample Division} \label{sec:sample}

To increase our sample, we combined these new data with the spectra of all S-complex asteroids studied by \cite{2024PSJ.....5..131S}. These include 55 objects classified as Q-, Sq-, S-, and Sr-types and the new Sx subclass introduced by \cite{2024PSJ.....5..131S}.  Thus, the total sample used in the present study is composed of 80 NEOs belonging to the S-complex. 

For the analysis that follows, the sample was divided into two and three subgroups based on their absolute magnitude. In the two-subgroup analysis, one subgroup is composed of 
47 NEOs with 20.0 $\leq$ $H$ $\leq$ 22.6 and the other of 33 NEOs with 22.9 $\leq$ $H$ $\leq$ 29.2. The range in $H$ values for these subgroups is the result of a 
compromise between trying to study the smallest possible objects and, at the same time, having enough of those objects for a statistically significant result. In the 
three-subgroup analysis, there are 41 NEOs with 20.0 $\leq$ $H$ $\leq$ 22.1, 24 NEOs with 22.2 $\leq$ $H$ $\leq$ 24.2 and 15 NEOs with 24.7 $\leq$ $H$ $\leq$ 29.2. Although dividing 
the sample into three subgroups has the disadvantage of increasing uncertainty, it can also help to further constrain $H$ values if compositional changes are found.

The approximate diameters of the NEOs were calculated using the equation of \cite{2007Icar..190..250P}. For those objects for which the geometric albedo is unknown (the 
majority), we assumed a value of 0.25 for Q- and Sq-types, and a value of 0.24 for S- and Sr-types \citep{2022AJ....163..165M}. A geometric albedo of 0.15 was assumed for NEOs 
classified as Sx-types, which is the value found for asteroid (52768) 1998 OR2 \citep{2022PSJ.....3..226B}. To our knowledge, this is the only NEO that exhibits characteristics 
consistent with an Sx-type for which the geometric albedo is known. The median diameter and median $H$ for each subgroup are presented in Table 2.

\begin{table}[htbp]
\caption{Fraction of Ordinary Chondrite Subtypes by Absolute Magnitude Range.} 
\label{tab:chondrite}
\begin{center}
\hspace{-2cm}
\begin{tabular}{lcccccc}
\toprule
$H$ range&Median $H$&Diameter Range&Median Diameter&H chondrite&L chondrite&LL chondrite  \\
& &(m)&(m)&(\%)&(\%)&(\%) \\
\midrule
\multicolumn{7}{c}{\textit{Two-subgroup analysis}} \\
\midrule
$20.0 \leq H \leq 22.6$ & 21.2 & 80--343  & 153 & 7.2$\pm$3.9  & 26.6$\pm$6.6 & 66.2$\pm$7.0 \\
$22.9 \leq H \leq 29.2$ & 24.2 & 4--70    & 39  & 10.4$\pm$5.5 & 48.4$\pm$8.6 & 41.2$\pm$8.5 \\
\midrule
\multicolumn{7}{c}{\textit{Three-subgroup analysis}} \\
\midrule
$20.0 \leq H \leq 22.1$ & 21.1 & 91--343  & 160 & 7.4$\pm$4.2  & 27.2$\pm$7.1 & 65.4$\pm$7.5 \\
$22.2 \leq H \leq 24.2$ & 23.2 & 39--97   & 60  & 7.9$\pm$5.7  & 33.2$\pm$9.5 & 58.9$\pm$9.9 \\
$24.7 \leq H \leq 29.2$ & 25.9 & 4--31    & 18  & 12.8$\pm$8.8 & 62.1$\pm$12.4 & 25.1$\pm$11.3 \\
\bottomrule
\end{tabular}
\end{center}
\end{table}

\section{Taxonomic Classification} \label{sec:tc}

The taxonomic classification of the NEOs was done using the Bus-DeMeo Taxonomy Classification Web tool\footnote{http://smass.mit.edu/busdemeoclass.html}. This online tool is an 
implementation of the Bus-DeMeo taxonomy \citep{2009Icar..202..160D}, which uses principal components analysis to classify asteroids among 25 different classes. In the original  
Bus-DeMeo taxonomy, the S-complex included five subclasses (S, Sa, Sq, Sr, Sv). For objects with high spectral slopes, the notation "w" was also added to their class. Q-types are considered an end-member class with spectral characteristics similar to S-types, but with deeper absorption bands and relatively flat spectral 
slopes. More recently, \cite{2024PSJ.....5..131S} defined a new subclass within the S-complex labeled Sx. These asteroids 
exhibit spectral characteristics consistent with objects in the S-complex, but with much weaker absorption bands, which could lead to an ambiguous classification in the C- 
and X-complexes. In the present work, we follow the same definition used by \cite{2024PSJ.....5..131S}, where both Q- and Sx-types are considered part of the 
S-complex. The taxonomic type for each of the new NEOs is 
presented in Table 3.

\begin{figure}[h]
\begin{center} 
\includegraphics[width=6cm,angle=0]{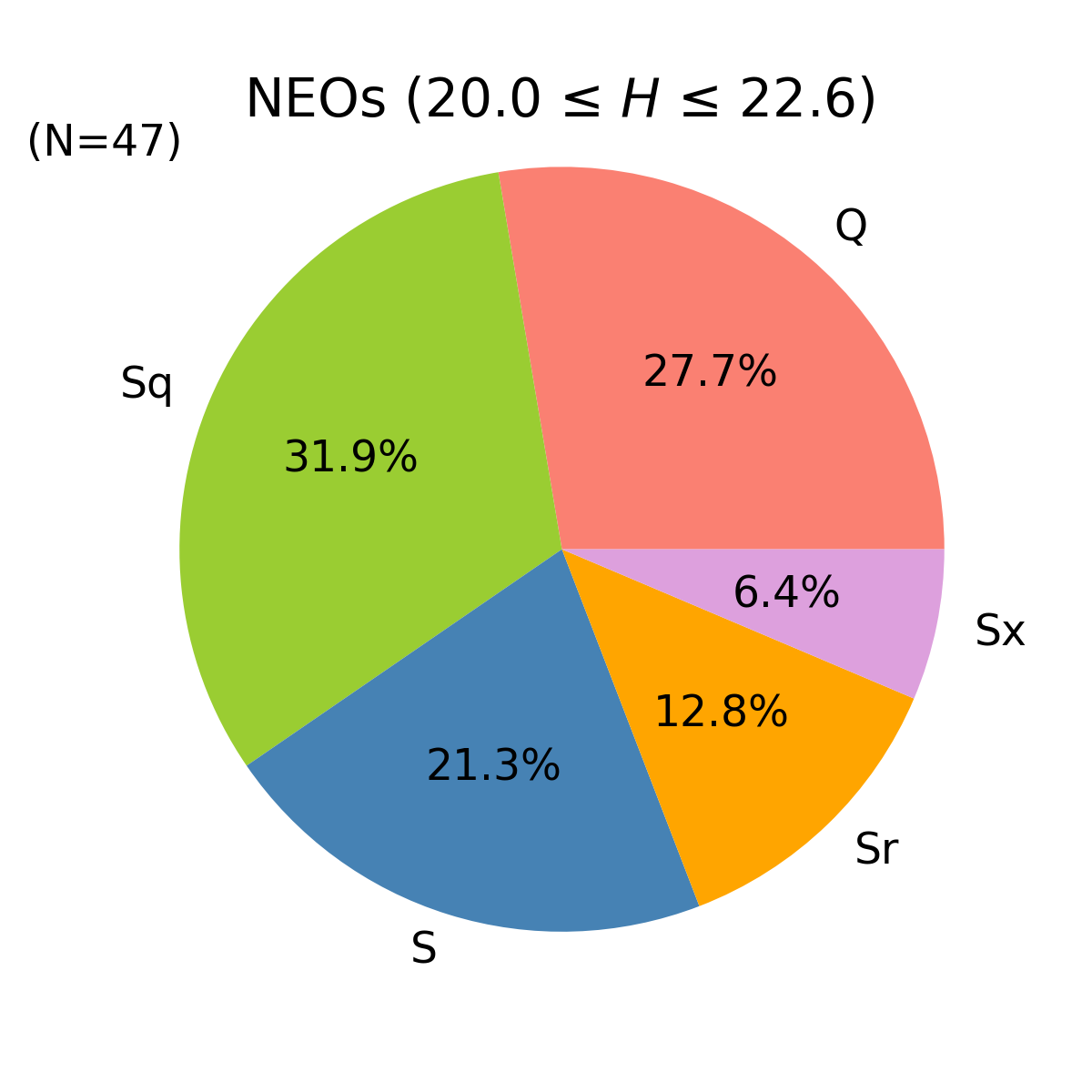}
\includegraphics[width=6cm,angle=0]{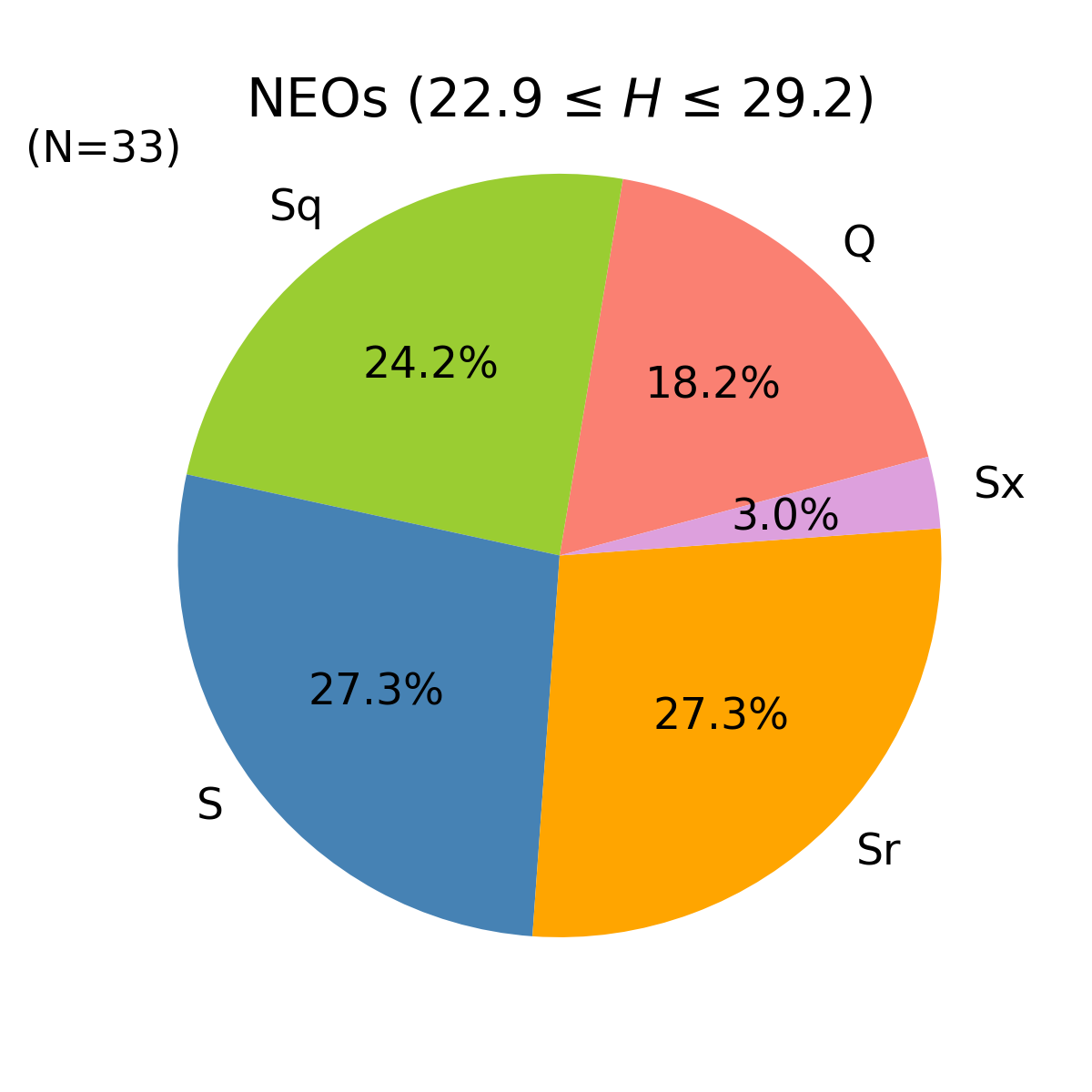}
\caption{Taxonomic distribution within the S-complex for the two-subgroup analysis. The number of objects for each subgroup is indicated.}
\label{f:Tax_pies}
\end{center}
\end{figure}

In the two-subgroup analysis (Figure \ref{f:Tax_pies}), we found that for the 
20.0 $\leq$ $H$ $\leq$ 22.6 subgroup, Sq-types represent the largest fraction (31.9\%), followed by Q-types (27.7\%). S-types make up 21.3\% of the sample and Sr- and Sx-types the remaining 19.2\%. In contrast, in the 22.9 $\leq$ $H$ $\leq$ 29.2 subgroup, S- and Sr-types represent the largest fraction (27.3\% each), Sq-types 
account for 24.2\% of the sample, and the rest (21.2\%) is comprised of Q- and Sx-types. Interestingly, Sa- and Sv-types are absent in both subgroups, which indicates that 
these classes are not only rare among the largest NEOs, but also among the smallest objects studied in this work. 

\begin{figure}
\begin{center} 
\includegraphics[width=6cm,angle=0]{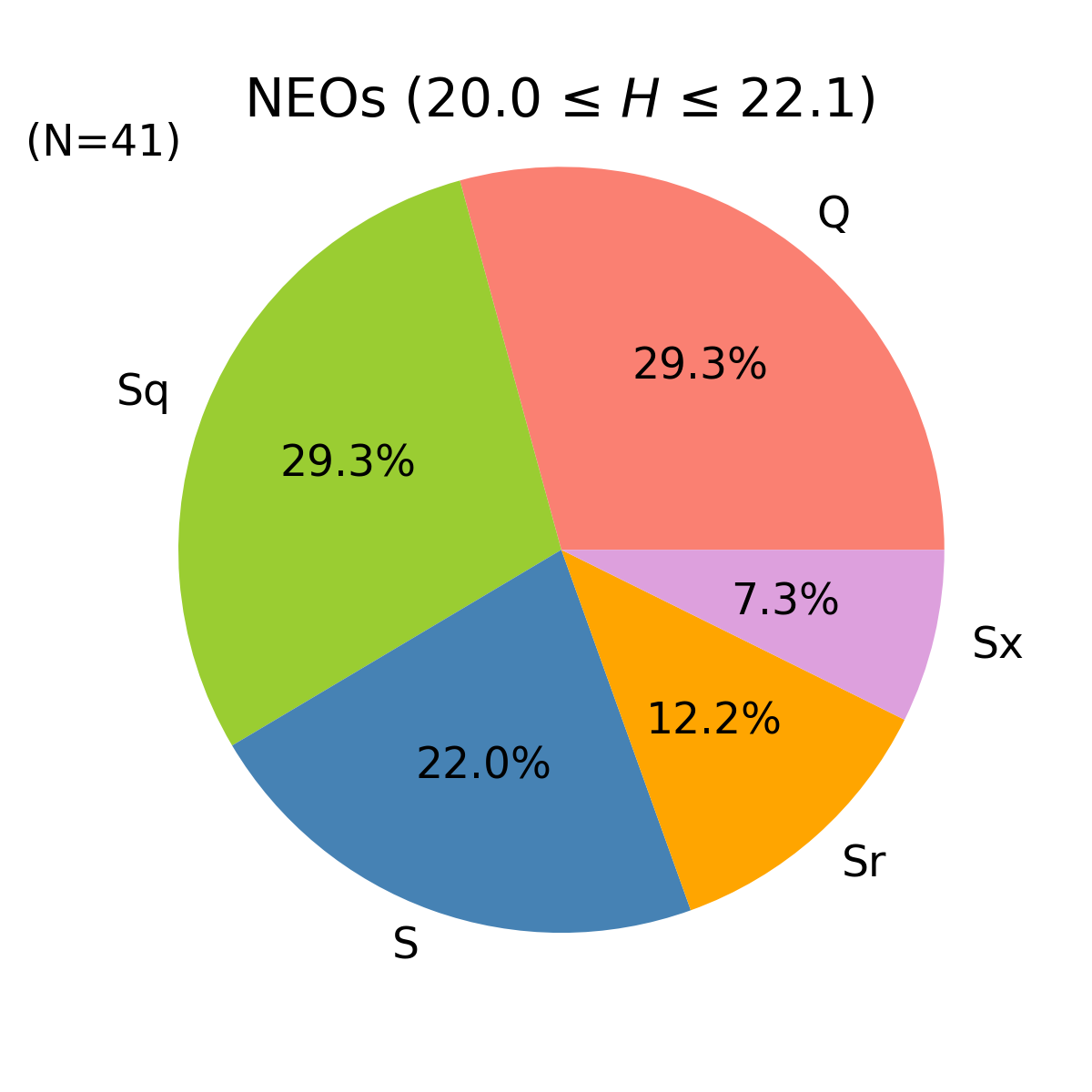}
\includegraphics[width=6cm,angle=0]{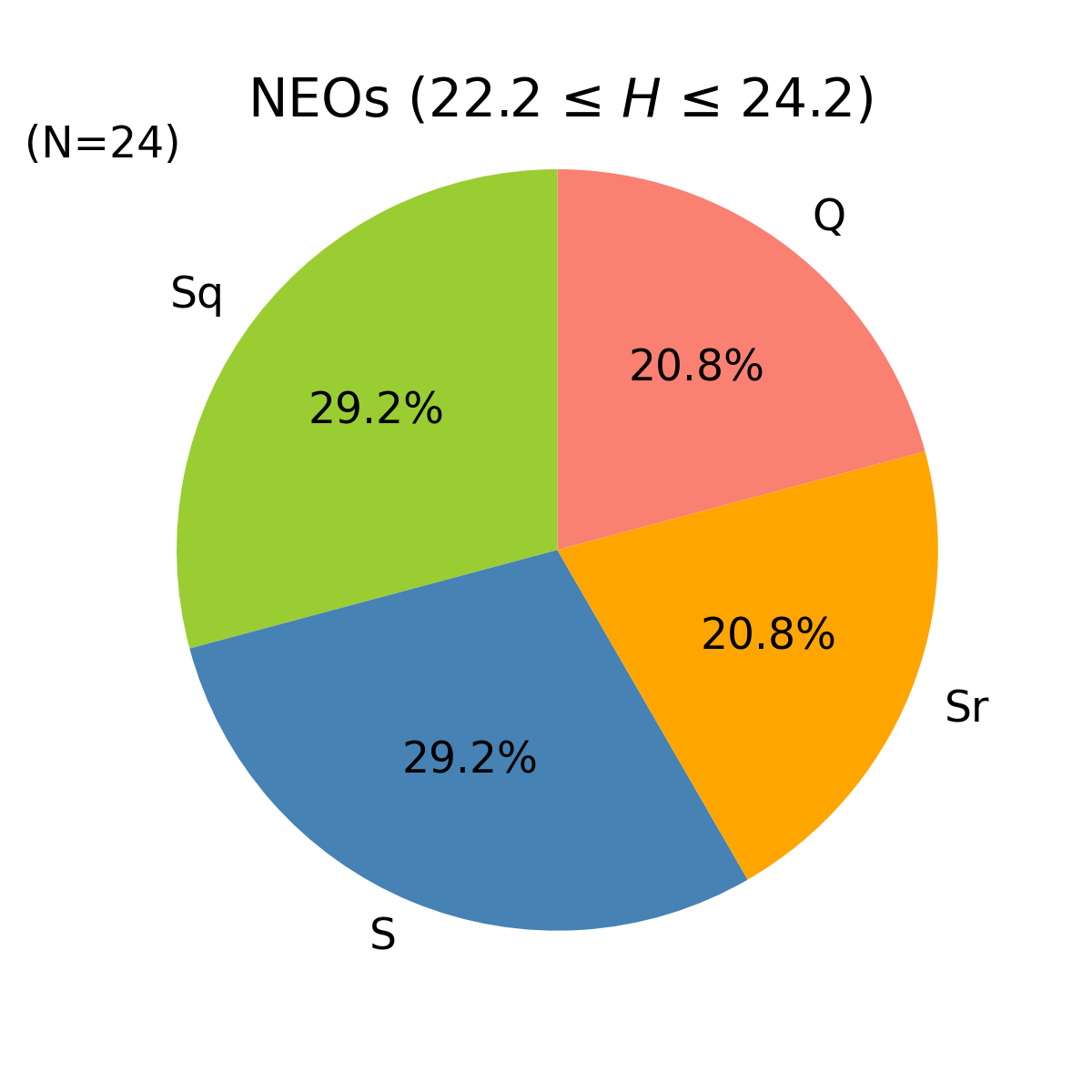}
\\
\includegraphics[width=6cm,angle=0]{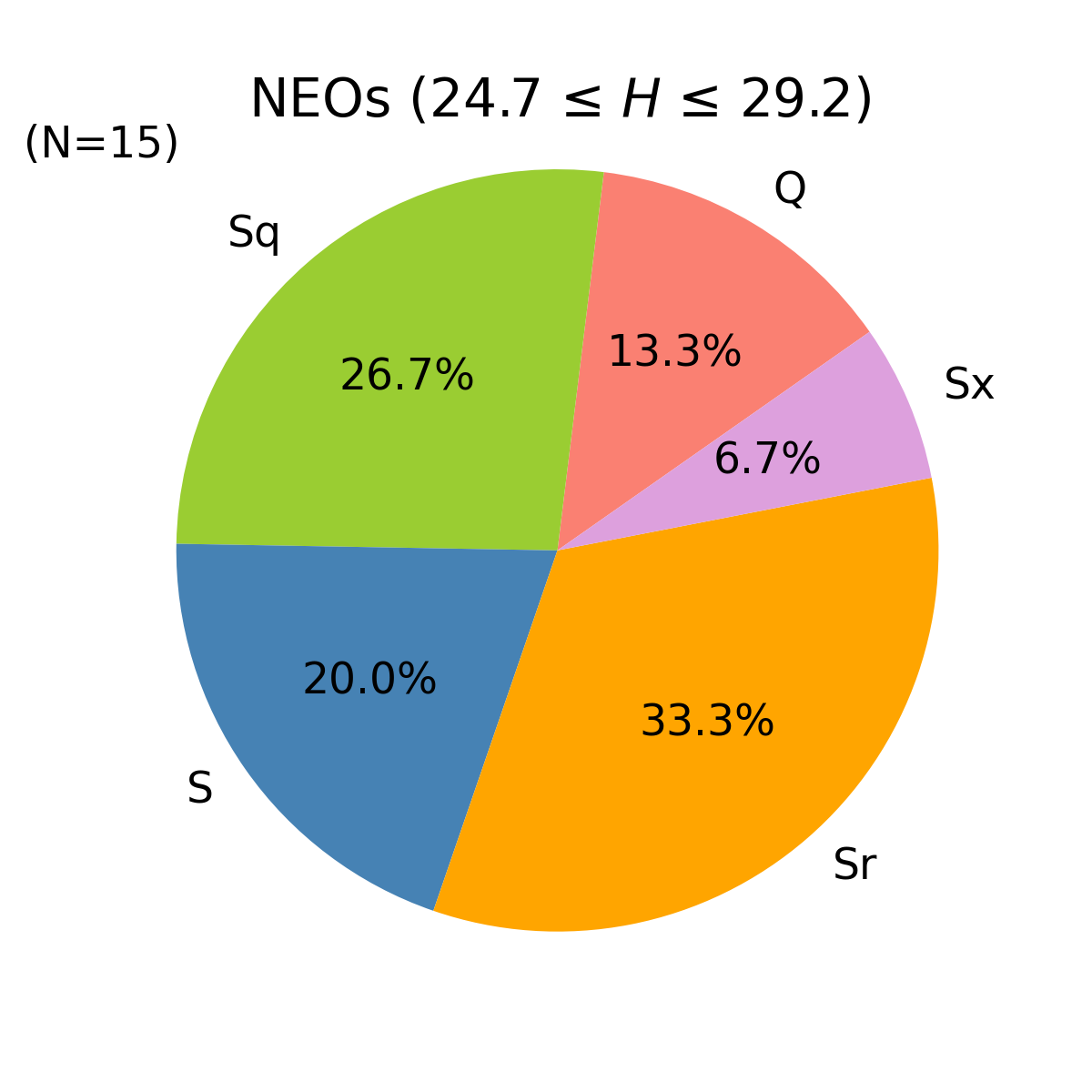}
\caption{Taxonomic distribution within the S-complex for the three-subgroup analysis. The number of objects for each subgroup is indicated.}
\label{f:Tax_pies_three_bins}
\end{center}
\end{figure}

In the three-subgroup analysis (Figure \ref{f:Tax_pies_three_bins}), we observe little change in 
the fraction of Sq-types across the three subgroups ($\sim$27-29\%), whereas the number of S-types was found to be higher in the 22.2 $\leq$ $H$ $\leq$ 24.2 
subgroup (29.2\%) compared to the 20.0 $\leq$ $H$ $\leq$ 22.1 subgroup (22\%) and the 24.7 $\leq$ $H$ $\leq$ 29.2 subgroup (20\%). A steady decrease in the amount of Q-types and 
an increase in the number of Sr-types are observed from the 20.0 $\leq$ $H$ $\leq$ 22.1 subgroup to the 24.7 $\leq$ $H$ $\leq$ 29.2 subgroup.

\startlongtable
\begin{deluxetable*}{cccc}

\tablecaption{\label{t:Table3}{\small Taxonomic Classification and Ordinary Chondrite Type for the Studied NEOs.}}
\tablewidth{0pt}
\tablehead{Number&Designation&Taxonomy&Ordinary Chondrite Type \\}
\startdata
462959&2011 DU&Q&LL \\
483508&2003 CR1&Q&LL \\
811221&2022 FR3&Qw&LL \\
826663&2021 PT&S&L \\
826936&1998 HH49&Q&LL \\
&2013 WV44&Q&LL \\
&2019 UT6&Qw&LL \\
&2020 UQ3&Sr&H \\
&2021 JG1&Srw&L \\
&2021 SZ4&Sq&L \\
&2023 PM&Srw&LL \\
&2023 QC&Sqw&L \\
&2023 UH5&Srw&L \\
&2023 UT&Srw&L \\
&2023 VD6&Q&LL \\
&2023 VR4&Sw&LL \\
&2024 BH&Q&LL \\
&2024 MT1&Sq&L \\
&2024 OM1&S&LL \\
&2024 QE2&Srw&L \\
&2024 WB&S&H \\
&2025 FU5&S&H \\
&2025 HM&S&L \\
&2025 OL1&S&L \\
&2025 OW&Sw&L \\
\enddata
\end{deluxetable*}

There are several factors that could be responsible for the different proportions of classes in the two studies; these include composition, space weathering, and grain 
size of the surface material. For equilibrated ordinary chondrites, \cite{2015Icar..252..129R} found that LL chondrites are typically classified (in descending order) as Q- 
and Sq-types, while L chondrites are more often classified as Sr-types and H chondrites are scattered among the S-complex. \cite{2022Icar..38014971D} obtained 
similar results for LL chondrites, but in the case of L chondrites they were found to be more common among Q-types and had similar proportions among S-, Sq- and 
Sr-types. H chondrites, on the other hand, showed a better spectral correlation with Sr- and Sv-types. Given this, compositional differences 
(e.g., fewer LL chondrites) may explain the decrease in Q-types and increase in Sr-types observed in both studies. Another possible explanation for the decrease of 
Q-types with increasing $H$ could be more objects being affected by space weathering, since Q-types are thought to have surfaces less 
space-weathered than Sq- and S-types \citep[e.g.,][]{2019Icar..324...41B}. However, this could only partially explain the results, since the number of Sq-types did not 
change much in the three-subgroup study, and there is no clear trend for the S-types. This interpretation is further complicated because 
changes in grain size can mask the spectral effects of space weathering, thereby affecting taxonomic classification \citep{2019PASJ...71..103H, 2023PSJ.....4...52B, 2024PSJ.....5..131S}. In the next section, we investigate in more detail the compositional distributions of these NEO groups.

%\newpage

\section{Compositional Analysis} \label{sec:comp}

The spectra of S-complex objects show absorption bands at $\sim$1 and 2 $\mu$m due to the presence of olivine and pyroxene. Spectral band parameters, including 
Band I and II centers and the Band Area Ratio (BAR), were measured from the spectra using a Python code following the procedure described in \cite{2020AJ....159..146S}. The linear continuum was fitted between the reflectance maxima at $\sim$0.74 and 1.4 $\mu$m, and at 
1.4 and 2.45 $\mu$m. After dividing out the linear continuum, band centers were measured  by fitting third- and fourth-order polynomials 
over the bottom of the absorption bands. Band areas are 
defined as the area between the linear continuum and the data curve and are calculated using trapezoidal numerical integration. The BAR was calculated 
as the ratio of the area of Band II to that of Band I. For a few objects, the 2 $\mu$m absorption band was too weak to be measured; as a result, the BAR was not calculated. A temperature correction derived by \cite{2012Icar..220...36S} was applied to the BAR in order to account 
for differences between the surface temperature of the NEOs and the room temperature at which spectral calibrations used for compositional analysis were 
obtained. The uncertainties are given by the standard deviation calculated from multiple measurements of each parameter. The spectral band parameters are reported in Table 4.

\startlongtable
\begin{deluxetable*}{cccccccccc}
\tablecaption{\label{t:Table4}{\small Spectral Band Parameters and Composition for the NEOs. The columns in this table are: object number and designation, Band I center (BIC), 
Band II center (BIIC), Band Area Ratio (BAR), molar contents of fayalite (Fa), and ferrosilite (Fs), and ol/(ol +px) ratio. The uncertainties for Fa, Fs, and the ol/(ol+px) ratio are 
from \cite{2020AJ....159..146S}; systematic differences between the methodologies used for compositional analysis may exceed these values.}}
\tablewidth{0pt}
\tablehead{Number&Designation&BIC ($\mu m$)&BIIC ($\mu m$)&BAR&Fa (mol$\%)$&Fs (mol$\%$)&ol/(ol+px) \\ }
\startdata
462959&2011 DU&0.979$\pm$0.002&1.942$\pm$0.020&0.46$\pm$0.02&28.8$\pm$2.0&24.2$\pm$1.4&0.61$\pm$0.04 \\
483508&2003 CR1&0.981$\pm$0.003&2.031$\pm$0.014&0.52$\pm$0.03&29.0$\pm$2.0&24.3$\pm$1.4&0.59$\pm$0.04 \\
811221&2022 FR3&1.013$\pm$0.004&1.977$\pm$0.017&0.40$\pm$0.01&30.6$\pm$2.0&25.4$\pm$1.4&0.62$\pm$0.04 \\
826663&2021 PT&0.949$\pm$0.006&1.922$\pm$0.021&1.24$\pm$0.13&24.8$\pm$2.0&21.4$\pm$1.4&0.42$\pm$0.04 \\
826936&1998 HH49&1.006$\pm$0.002&2.083$\pm$0.006&0.45$\pm$0.03&30.5$\pm$2.0&25.3$\pm$1.4&0.61$\pm$0.04 \\
&2013 WV44&0.999$\pm$0.003&1.977$\pm$0.017&0.40$\pm$0.05&30.3$\pm$2.0&25.2$\pm$1.4&0.62$\pm$0.04 \\
&2019 UT6&1.000$\pm$0.003&2.015$\pm$0.012&0.51$\pm$0.06&30.3$\pm$2.0&25.2$\pm$1.4&0.60$\pm$0.04 \\
&2020 UQ3&0.936$\pm$0.002&1.940$\pm$0.015&1.40$\pm$0.11&22.3$\pm$2.0&19.7$\pm$1.4&0.39$\pm$0.04  \\
&2021 JG1&0.956$\pm$0.003&1.938$\pm$0.011&0.60$\pm$0.05&25.9$\pm$2.0&22.2$\pm$1.4&0.58$\pm$0.04 \\
&2021 SZ4&0.943$\pm$0.001&1.960$\pm$0.007&0.68$\pm$0.02&23.7$\pm$2.0&20.6$\pm$1.4&0.56$\pm$0.04 \\
&2023 PM&0.970$\pm$0.010&-&-&27.9$\pm$2.0&23.5$\pm$1.4&- \\
&2023 QC&0.951$\pm$0.002&1.932$\pm$0.010&0.62$\pm$0.05&25.1$\pm$2.0&21.6$\pm$1.4&0.57$\pm$0.04 \\
&2023 UH5&0.943$\pm$0.003&1.999$\pm$0.017&0.67$\pm$0.05&23.7$\pm$2.0&20.6$\pm$1.4&0.56$\pm$0.04 \\
&2023 UT&0.962$\pm$0.004&1.934$\pm$0.021&0.61$\pm$0.04&26.8$\pm$2.0&22.8$\pm$1.4&0.57$\pm$0.04 \\
&2023 VD6&0.988$\pm$0.005&2.016$\pm$0.018&0.39$\pm$0.04&29.6$\pm$2.0&24.7$\pm$1.4&0.62$\pm$0.04 \\
&2023 VR4&0.960$\pm$0.010&-&-&26.5$\pm$2.0&22.6$\pm$1.4&- \\
&2024 BH&1.010$\pm$0.002&-&-&30.6$\pm$2.0&25.4$\pm$1.4&- \\
&2024 MT1&0.954$\pm$0.001&1.954$\pm$0.005&0.89$\pm$0.03&25.6$\pm$2.0&22.0$\pm$1.4&0.51$\pm$0.04 \\
&2024 OM1&1.014$\pm$0.002&2.041$\pm$0.019&0.69$\pm$0.06&30.6$\pm$2.0&25.4$\pm$1.4&0.55$\pm$0.04 \\
&2024 QE2&0.951$\pm$0.004&1.969$\pm$0.010&0.75$\pm$0.03&25.1$\pm$2.0&21.6$\pm$1.4&0.54$\pm$0.04 \\
&2024 WB&0.921$\pm$0.003&1.893$\pm$0.042&0.91$\pm$0.01&18.9$\pm$2.0&17.3$\pm$1.4&0.50$\pm$0.04 \\
&2025 FU5&0.926$\pm$0.003&1.956$\pm$0.007&1.05$\pm$0.06&20.1$\pm$2.0&18.1$\pm$1.4&0.47$\pm$0.04 \\
&2025 HM&0.934$\pm$0.004&1.965$\pm$0.023&0.91$\pm$0.04&21.9$\pm$2.0&19.4$\pm$1.4&0.50$\pm$0.04 \\
&2025 OL1&0.930$\pm$0.001&-&-&21.0$\pm$2.0&18.8$\pm$1.4&- \\
&2025 OW&0.948$\pm$0.002&1.981$\pm$0.014&0.76$\pm$0.06&24.6$\pm$2.0&21.3$\pm$1.4&0.54$\pm$0.04 \\
\enddata
\end{deluxetable*}

\begin{figure*}[!ht]
\begin{center}
\includegraphics[height=8.5cm]{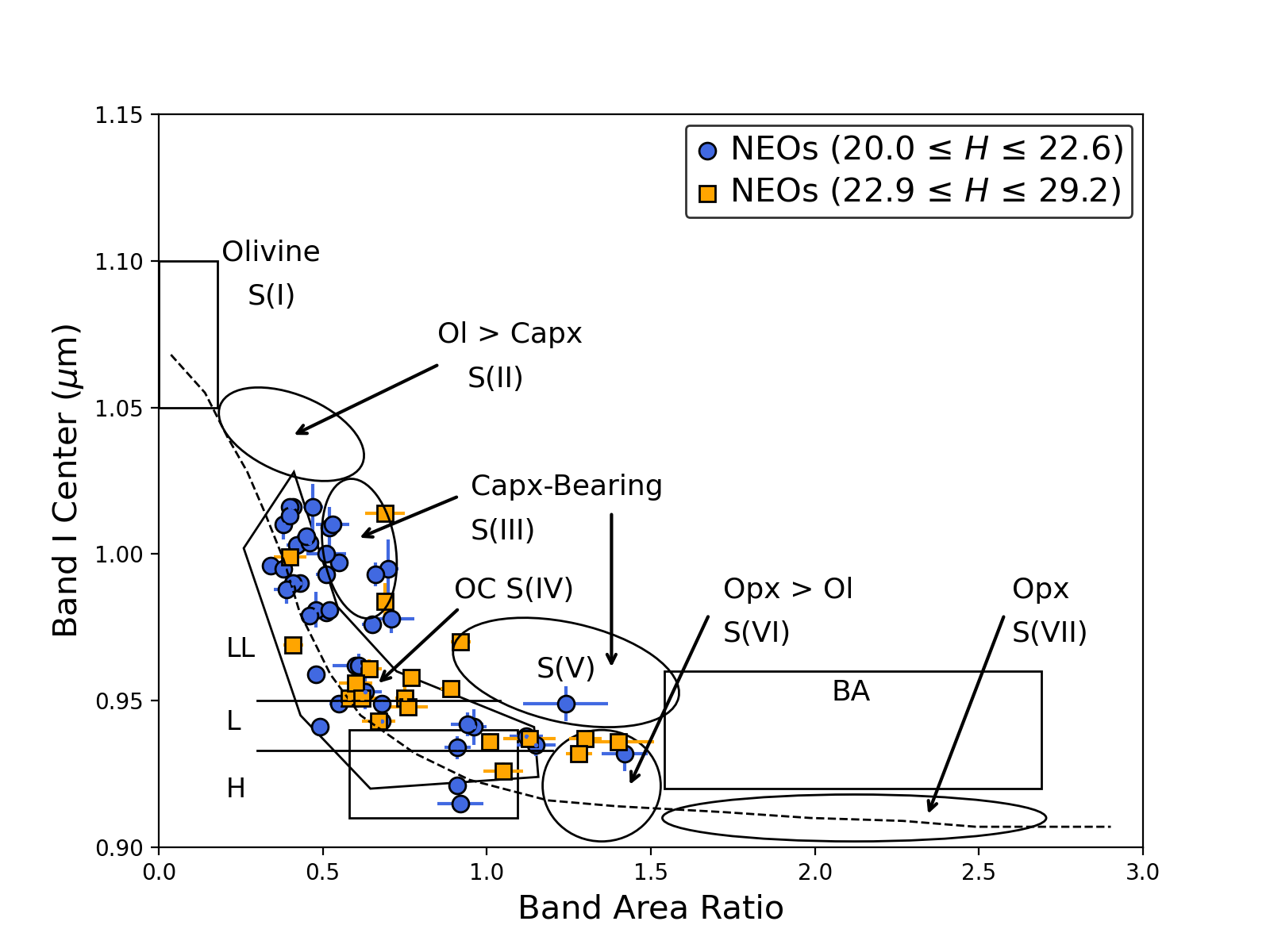}
\caption{\label{f:BIC_BAR}{\small Band I center vs. BAR diagram from \cite{1993Icar..106..573G} for the two-subgroup analysis. The polygonal region corresponds to 
the S(IV) subtype associated with ordinary chondrites (OC). The horizontal lines represent the approximate boundaries for ordinary chondrites found 
by \cite{2020AJ....159..146S}. The rectangular zone overlapping the S(IV) subtype represents the spectral zone for acapulcoite-lodranite clan meteorites found 
by \cite{2019M&PS...54..157L}. The rectangular zone (BA) includes the pyroxene-dominated basaltic achondrite assemblages. The dashed curve indicates the location of the 
olivine-orthopyroxene mixing line \citep{1986JGR....9111641C}.}}
\end{center}
\end{figure*}

Figure \ref{f:BIC_BAR} shows the Band I center versus BAR diagram for the two-subgroup analysis. Objects whose spectra 
were obtained with the 0.8 $\mu$m dichroic are not included because their BAR would appear shifted to the right due to the effect of truncating 
the spectra at this wavelength \citep{2020AJ....159..146S}. In general, we see that most objects in the 
20.0 $\leq$ $H$ $\leq$ 22.6 subgroup are clustered in the upper part of the S(IV) subtype, where LL chondrites normally fall, while smaller objects in the 22.9 $\leq$ $H$ $\leq$ 29.2 subgroup are 
mostly located in the middle to lower part, where L and H chondrites are typically found. The presence of calcic pyroxene could explain objects classified as S(III) and S(V) 
subtypes, while those falling in the S(VI) subtype might contain low-Ca pyroxene and an olivine-to-pyroxene (ol/(ol+px)) ratio lower than that of ordinary chondrites. However,  it is also important 
to keep in mind that a low signal-to-noise ratio in the 2 $\mu$m band can result in an overestimation of the BAR, and consequently, a lower ol/(ol+px) ratio for some of these 
objects \citep{2024PSJ.....5..131S}.

\begin{figure*}
\begin{center}
\includegraphics[height=8.5cm]{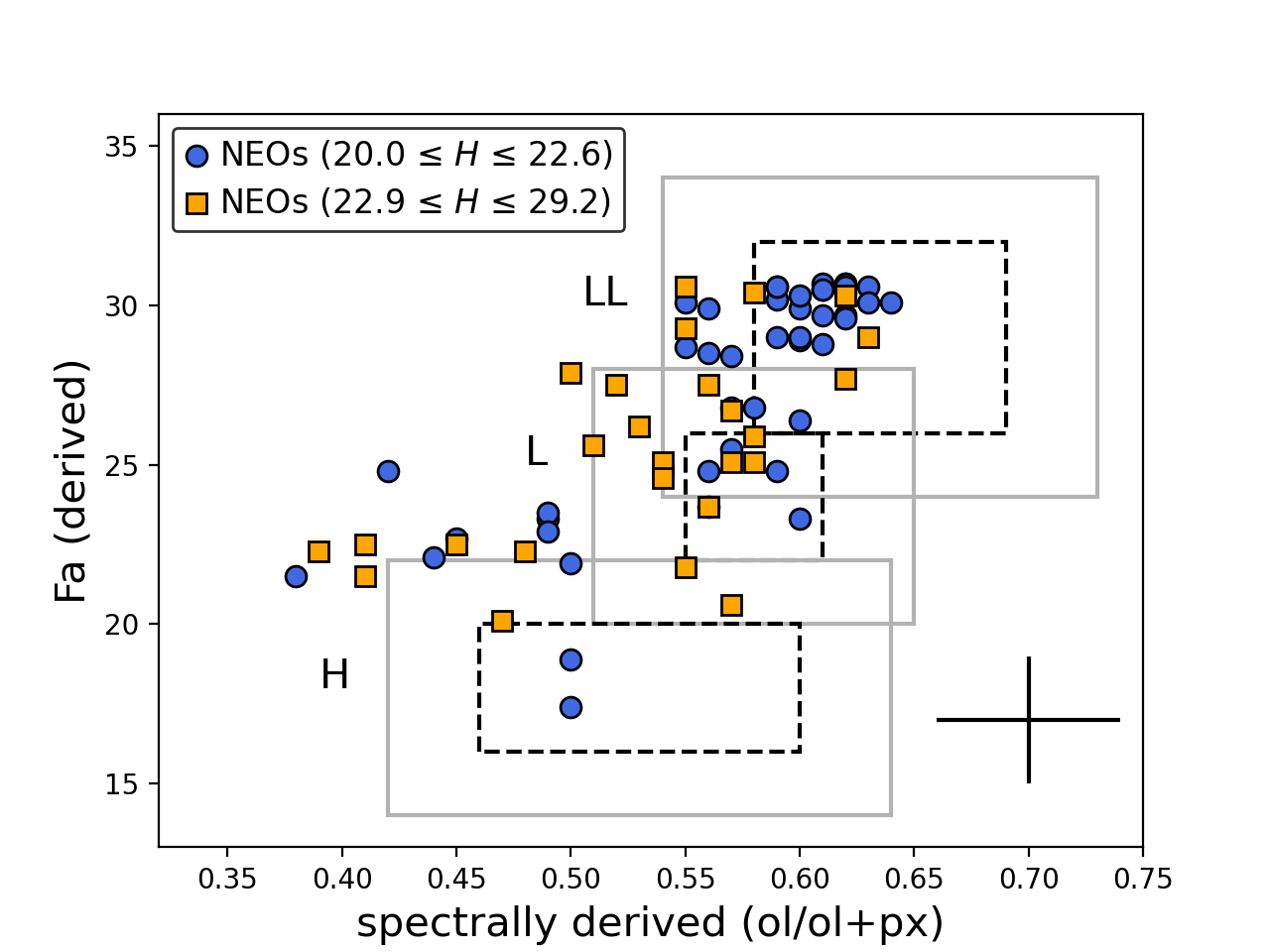}

\caption{{\small Molar content of fayalite (Fa) vs. ol/(ol+px) ratio derived for the two-subgroup analysis. Black dashed boxes
represent the range of measured values for each ordinary chondrite subtype. Gray solid boxes represent the uncertainties associated with the spectrally derived
values \citep{2020AJ....159..146S} . The error bars correspond to the uncertainties derived by \cite{2020AJ....159..146S}, 2.0 mol\% for Fa and 0.04 for the ol/(ol+px) ratio. 
Figure adapted from \cite{2010Icar..208..789D}.}}
\label{f:Fa_olopx}
\end{center}
\end{figure*}

The composition of the NEOs was determined using the equations derived by \cite{2020AJ....159..146S}. The 
Band I center was used to calculate the olivine and pyroxene chemistries, which are given by the molar percentage of fayalite (Fa) and ferrosilite (Fs), respectively. For those objects where the 
BAR was measured, the ol/(ol+px) ratio was also calculated (Table 4). These values were combined with those obtained by \cite{2024PSJ.....5..131S} for 
the previously observed NEOs. The Fa values versus the ol/(ol+px) ratios for the two-subgroup analysis are shown in Figure \ref{f:Fa_olopx}. It can be seen that the majority of 
objects in the 20.0 $\leq$ $H$ $\leq$ 22.6 subgroup fall in the LL chondrite region, while those in the 22.9 $\leq$ $H$ $\leq$ 29.2 subgroup show a greater affinity for L chondrites.

The next step in our analysis is to calculate the probability for each object of being classified as an H, L, or LL chondrite. For this, we used the same procedure described in
\cite{2024PSJ.....5..131S}, which consists of building a supervised machine learning algorithm based on multinomial logistic regression. This algorithm 
is used for multiclass classification and yields a probability distribution function of ordinary chondrite subtypes for the NEOs. Multinomial logistic 
regression uses the softmax function to convert a vector of $K$ real numbers into a probability distribution of $K$ possible outcomes. Given an input feature $x^{(i)}$ and a class label $y^{(i)}$ $\in$ \{1, 2, ..., $K$\}, where $K$ is the number of classes, we want to estimate the probability that $y^{(i)}$ equals $k$ for each value of $k$=1, 2, …, $K$. The predicted probability for the $kth$ class is given by

\begin{equation}
P(y^{(i)}=k\vert x^{(i)}; \theta)= \frac{e^{{ \theta^{(k)}}^T  x^{(i)}}}{\sum_{j=1}^{K} e^{ {\theta^{(j)}}^T  x^{(i)}}}
\end{equation}

where ${\theta^Tx}$ is the regression result, i.e., the sum of the variables weighted by the coefficients. The difference between the actual and the predicted 
values gives the error of the model, which is calculated by the cost function.

\begin{figure}[h]
\hspace{-5.0mm}
\includegraphics[width=9.5cm,angle=0]{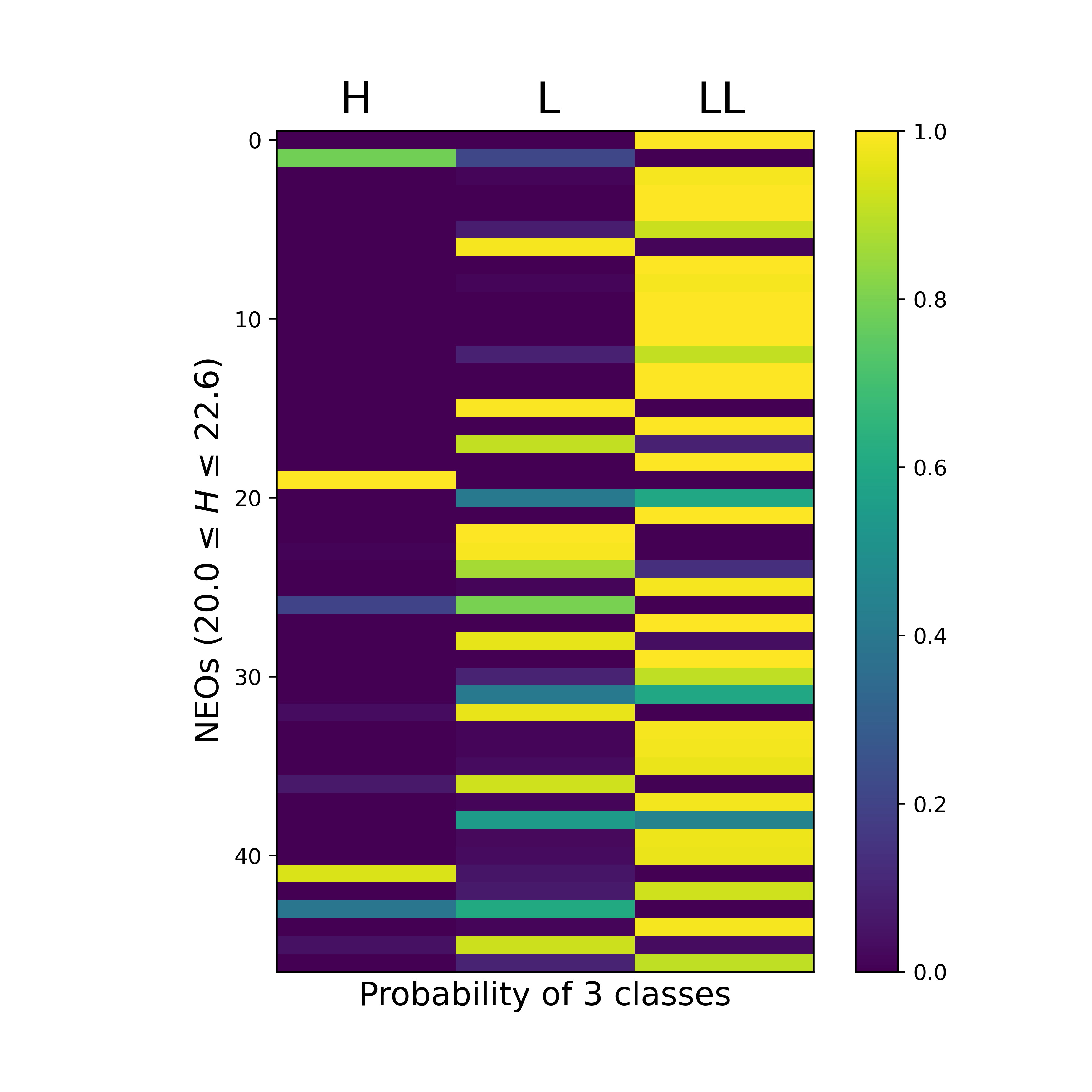}
\hspace{-11.0mm}
\includegraphics[width=9.5cm,angle=0]{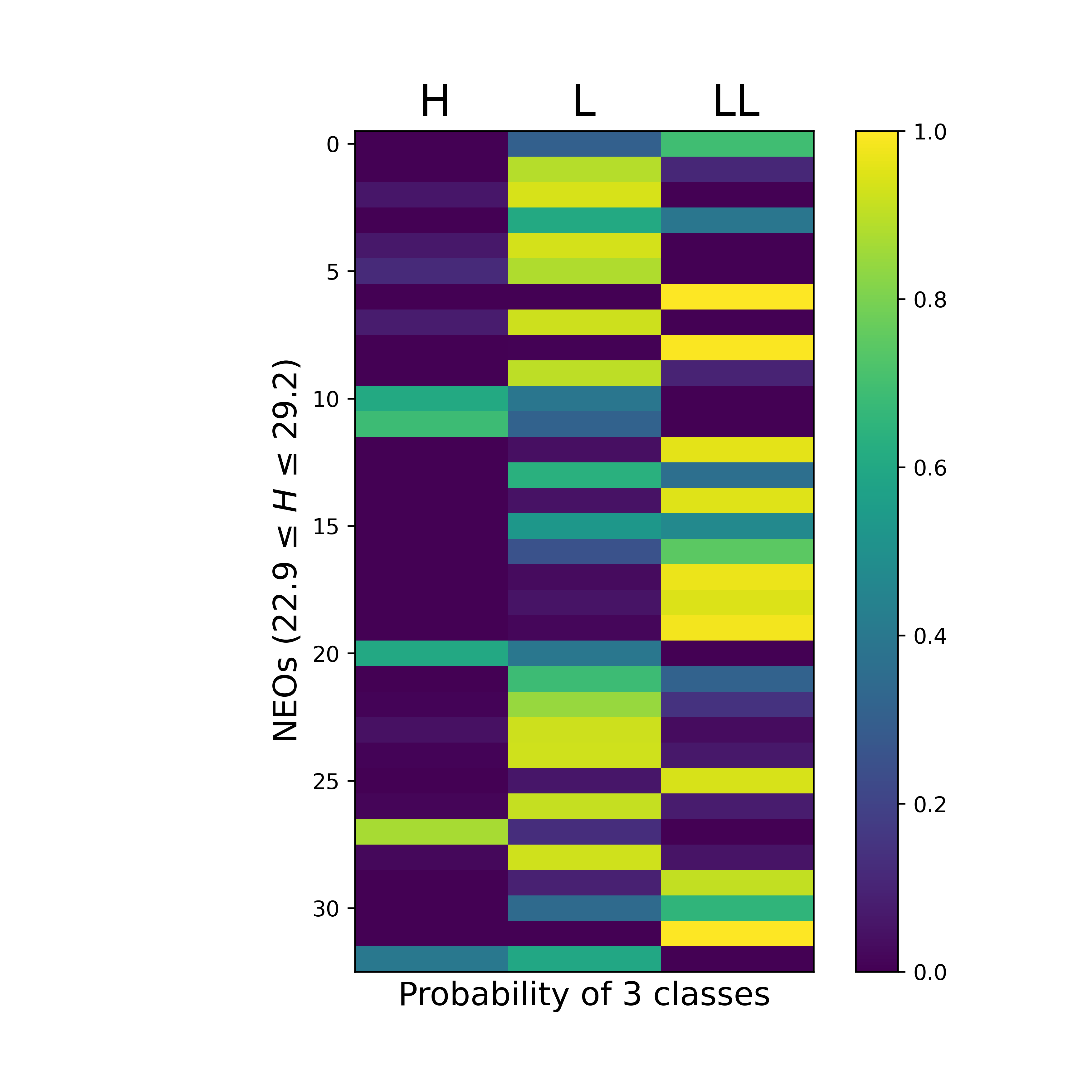}
\caption{Probability distribution functions of ordinary chondrite subtypes for NEOs in the two-subgroup analysis. Each row represents an NEO and each column represents
an ordinary chondrite subtype (H, L, LL).}
\label{f:prob_dist_functs}
\end{figure}

In order to train and test the model, a synthetic multiclass dataset was created. For each ordinary chondrite subtype, 500 data points with three input 
variables (ol/(ol+px), Fa, Fs) were generated from a uniform distribution using the boundaries found by \cite{2020AJ....159..146S}. This dataset was then split into a training and 
a testing set. For the testing set, 20\% of the whole sample was randomly selected. To prevent overfitting, we employed regularization; i.e, we used the L2 (ridge) penalty, which is a type of regularization that adds a penalty term equal to the sum of the squares of the weights to the 
cost function. The strength of the penalty is controlled by the hyperparameter C (the inverse of the regularization strength). The model was fit to the 
training set using the Limited-memory Broyden-Fletcher-Goldfarb-Shanno optimization algorithm with a tenfold cross-validation. A grid search method was used to find the optimal value of C. The performance of the model was evaluated on the testing set and the F1 score and logarithmic loss (log 
loss) were calculated. Figure \ref{f:prob_dist_functs} shows the probability 
distribution functions for NEOs in the two-subgroup analysis. Each row represents an NEO and each column represents an ordinary chondrite 
subtype (H, L, LL). The classification of the objects is included in Table 3.

\begin{figure}[b]
\begin{center}
\includegraphics[height=9.0cm]{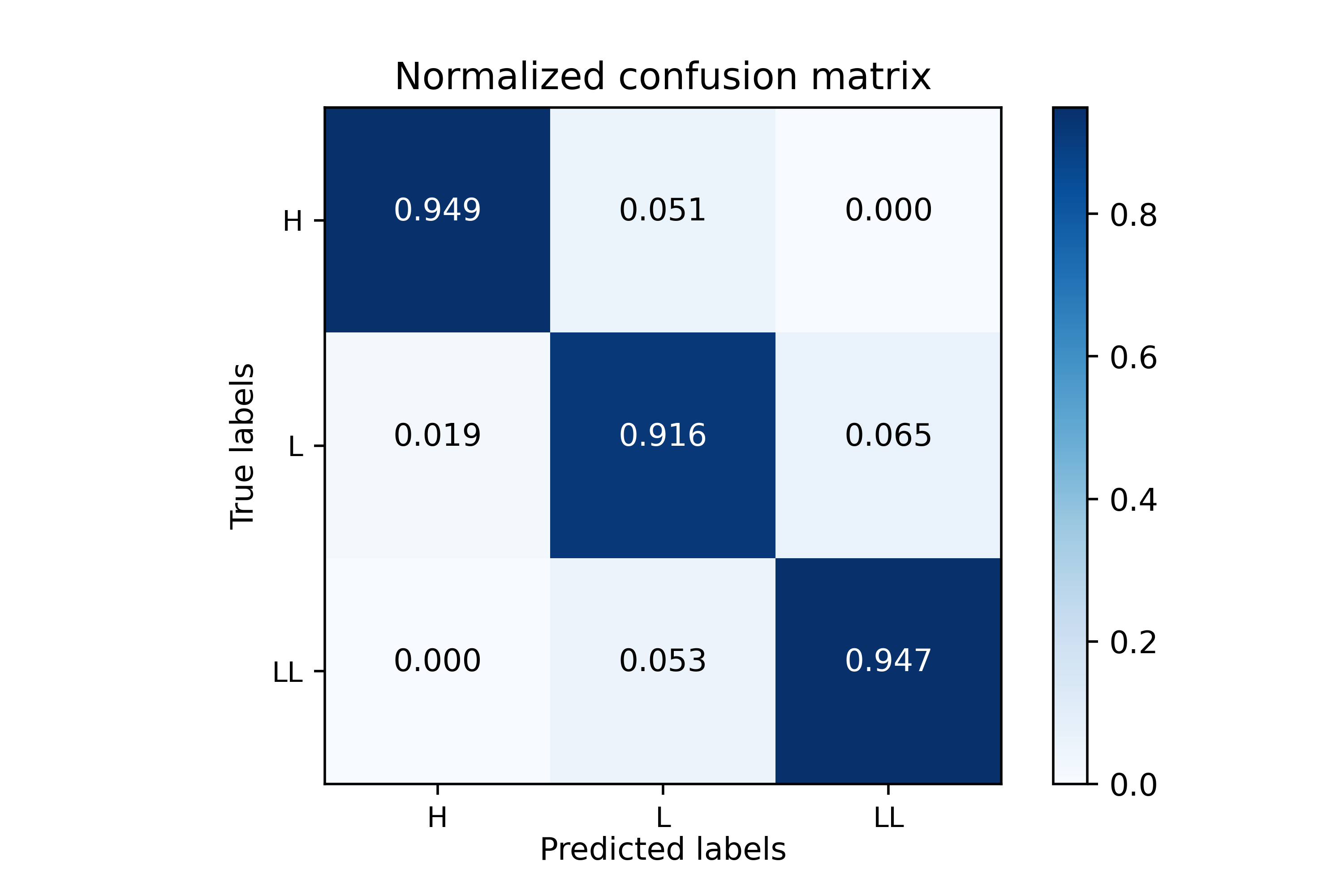}
\caption{\label{f:conf_matrix}{\small Normalized confusion matrix resulting from the multinomial logistic regression model. Labels correspond to the three 
ordinary chondrite subtypes (H, L, LL).}}
\end{center}
\end{figure}

From the multinomial logistic regression model, we found that 95\% of the H chondrites were 
correctly classified and 5\% were classified as L chondrites. Approximately 92\% of the L chondrites were assigned the correct label and the 
remaining 8\% were classified as H and LL chondrites. For the LL chondrites, 95\% were correctly classified by the model and the rest were classified 
as L chondrites (Figure \ref{f:conf_matrix}). Because the classifier is not perfect, the observed fractions ($P_{obs}$) are a biased version of the true fractions ($P_{true}$). The confusion matrix 
$M$ (where entry $M[i,j]$ is the probability that a true class-$i$ object is predicted as class-$j$) relates the two via a linear system $P_{obs}$ = $M$$^{T}$$\cdot$ $P_{true}$. 
Inverting this system yields the corrected bias-free fractions $P_{true}$ that are reported in this study. On average, the corrected fractions differ by only 1\% from the observed fractions. Mean $P_{true}$ values for the ordinary chondrite subtypes are reported in Table 2.

Two distinct sources of uncertainty are considered when calculating the mean fractions for the ordinary chondrite subtypes. The first one is the sampling uncertainty 
on a proportion (statistical), i.e., the uncertainty that comes purely from having a finite sample of N asteroids. This statistical uncertainty can be estimated using the 
Wilson score interval at 1$\sigma$. This is a robust method for calculating binomial proportion confidence intervals, offering accurate error estimation for small sample 
sizes and edge probabilities (near 0 or 1). The second source of uncertainty is the misclassification uncertainty (systematic), which can be seen in the confusion matrix. 
Each entry of $M$ is itself estimated from a finite test set, so it carries its own uncertainty $\sigma_{ij}$ (again from the Wilson interval). To propagate this into an 
uncertainty on $P_{true}$, each nonzero cell $M[i,j]$ is perturbed by $\pm$$\sigma_{ij}$, the system is re-solved, and the resulting changes in $P_{true}$ are combined 
in quadrature. The statistical and systematic contributions are independent, so they are combined in quadrature to give the total 1$\sigma$ uncertainty. The systematic uncertainty was found to be very small, $\sim$0.2-1.0\% across all classes and both groups. This means that the classifier is good enough (92\%-95\% accuracy) that misclassification introduces only a minor systematic bias. Thus, the dominant uncertainty is statistical, coming from the small sample sizes.

\begin{figure}[b] 
\begin{center}
\includegraphics[width=7cm,angle=0]{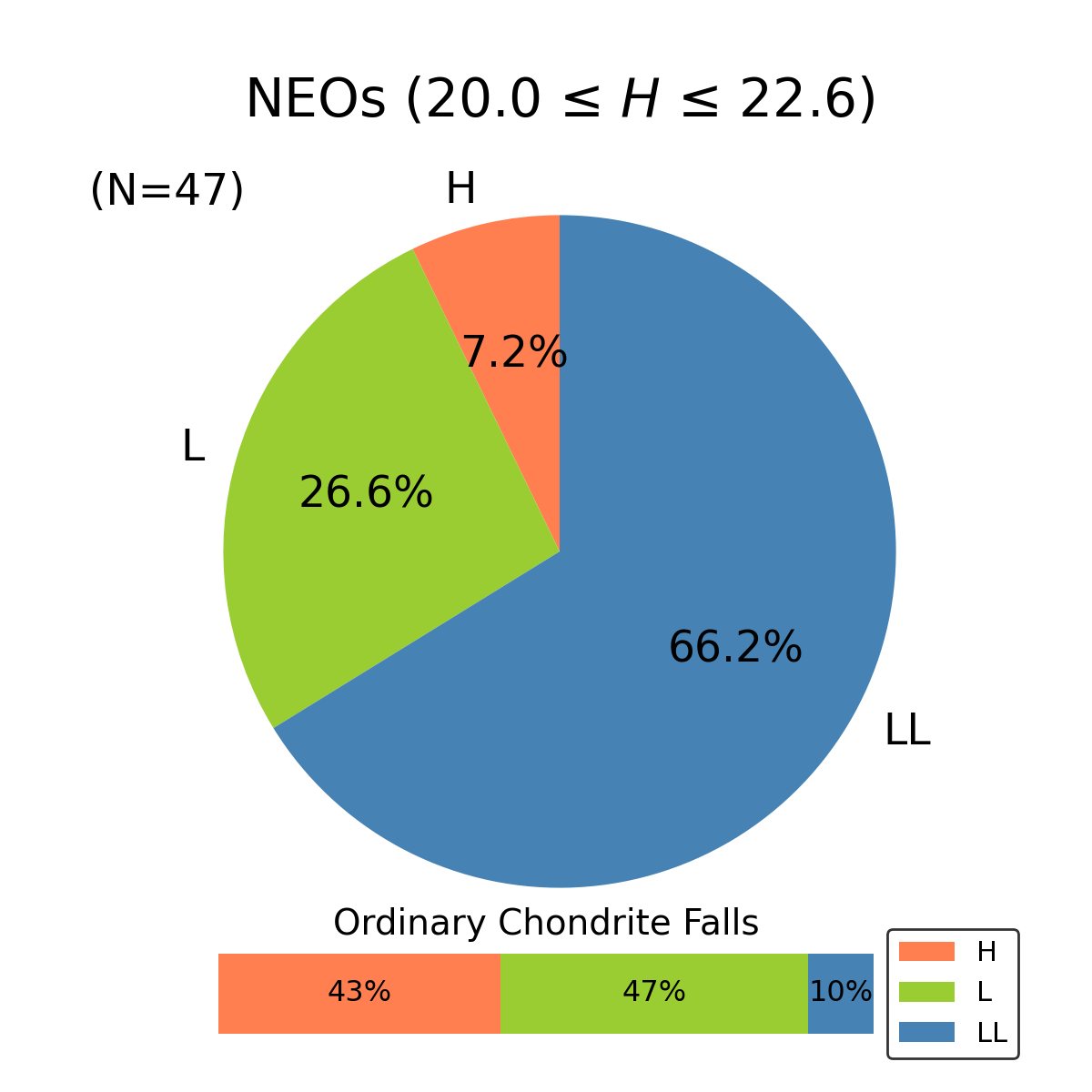}
\includegraphics[width=7cm,angle=0]{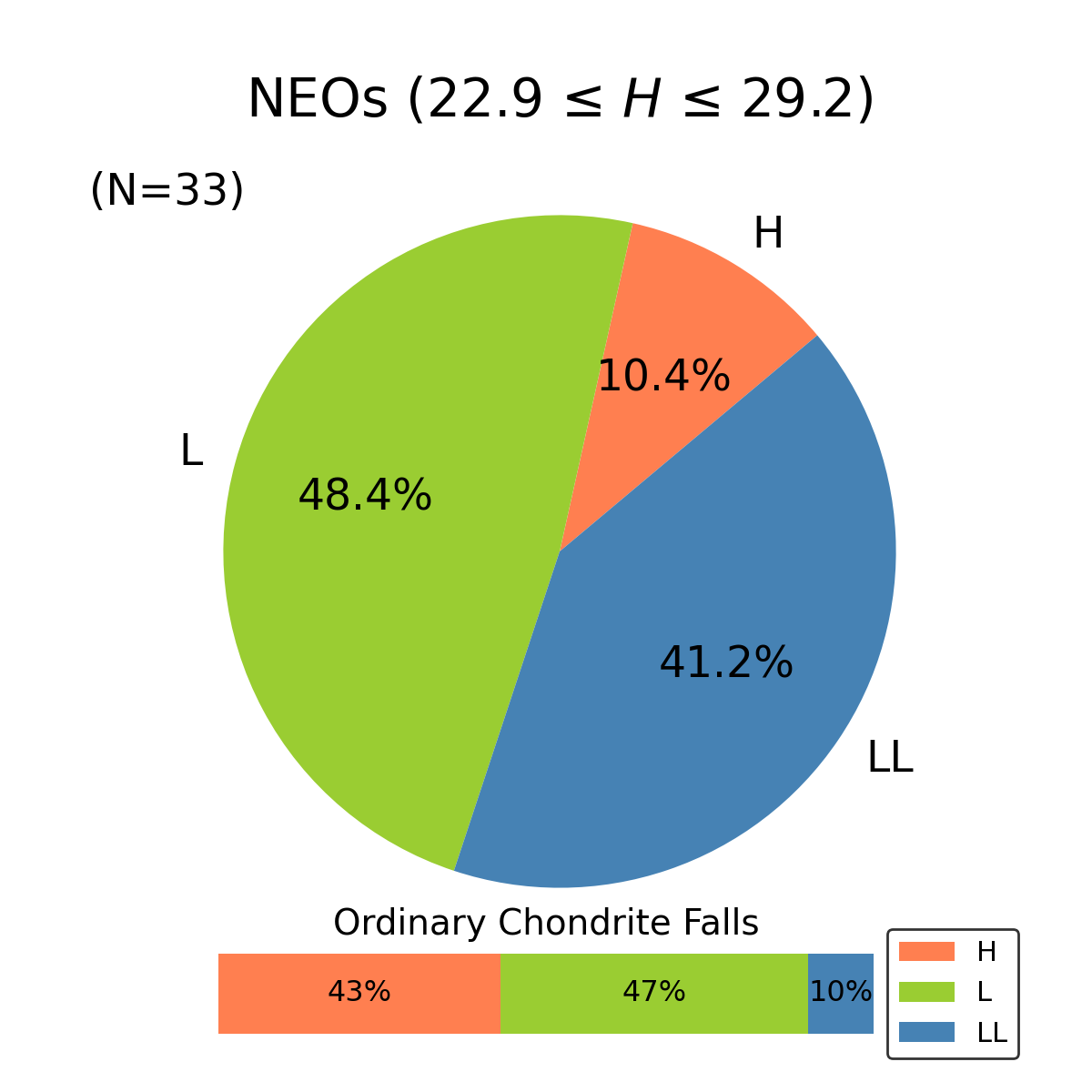}
\caption{Mean values for the ordinary chondrite subtypes for the two-subgroup analysis. The number of objects for each subgroup is indicated. The fractions of ordinary chondrite falls are shown in the horizontal bar.}
\label{f:prob_pie_charts}
\end{center}
\end{figure}

As can be seen in Figure \ref{f:prob_pie_charts}, the fraction of objects with H chondrite-like compositions only shows a modest increase, rising from 
7.2$\pm$3.9\% in the 20.0 $\leq$ $H$ $\leq$ 22.6 subgroup to 10.4$\pm$5.5\% in the 22.9 $\leq$ $H$ $\leq$ 29.2 subgroup. The subgroup with the smallest NEOs (22.9 $\leq$ $H$ $\leq$ 29.2) shows an increase in the number of objects with 
L chondrite-like compositions (48.4$\pm$8.6\%) compared to the 20.0 $\leq$ $H$ $\leq$ 22.6 subgroup (26.6$\pm$6.6\%), while the fraction of NEOs with LL chondrite-like compositions decreased 
from 66.2$\pm$7.0\% in the 20.0 $\leq$ $H$ $\leq$ 22.6 subgroup to 41.2$\pm$8.5\% in the 22.9 $\leq$ $H$ $\leq$ 29.2 subgroup. 

The results from the two-subgroup analysis confirm the size dependence in the composition of S-complex NEOs and provide a starting point for constraining the sizes of ordinary chondrite parent bodies. For NEOs in the 22.9 $\leq$ $H$ $\leq$ 29.2 subgroup (size range of $\sim$4-70 m), the fraction of H 
chondrite-like objects is too low compared to the H chondrite falls (43\%), suggesting that the parent bodies of these meteorites are likely below this size 
range. Moreover, if some of them originated from Koronis subclusters and occupy orbits with high semimajor axes, it is possible that they could be observationally underrepresented in our sample. As a conservative value, we establish an upper size limit of 39 m (the median of this subgroup) for these bodies.

\begin{figure}
\begin{center} 
\includegraphics[width=7cm,angle=0]{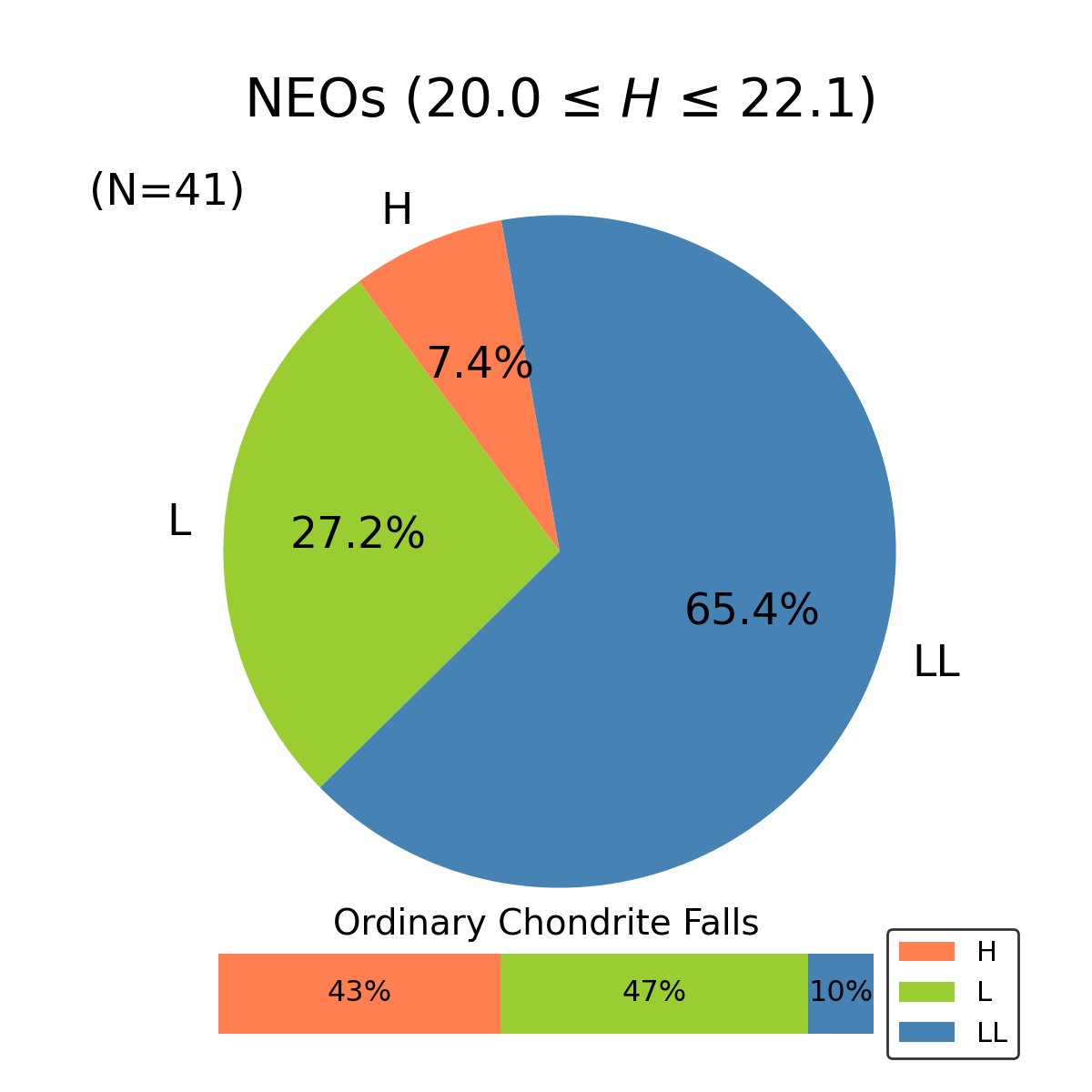}
\includegraphics[width=7cm,angle=0]{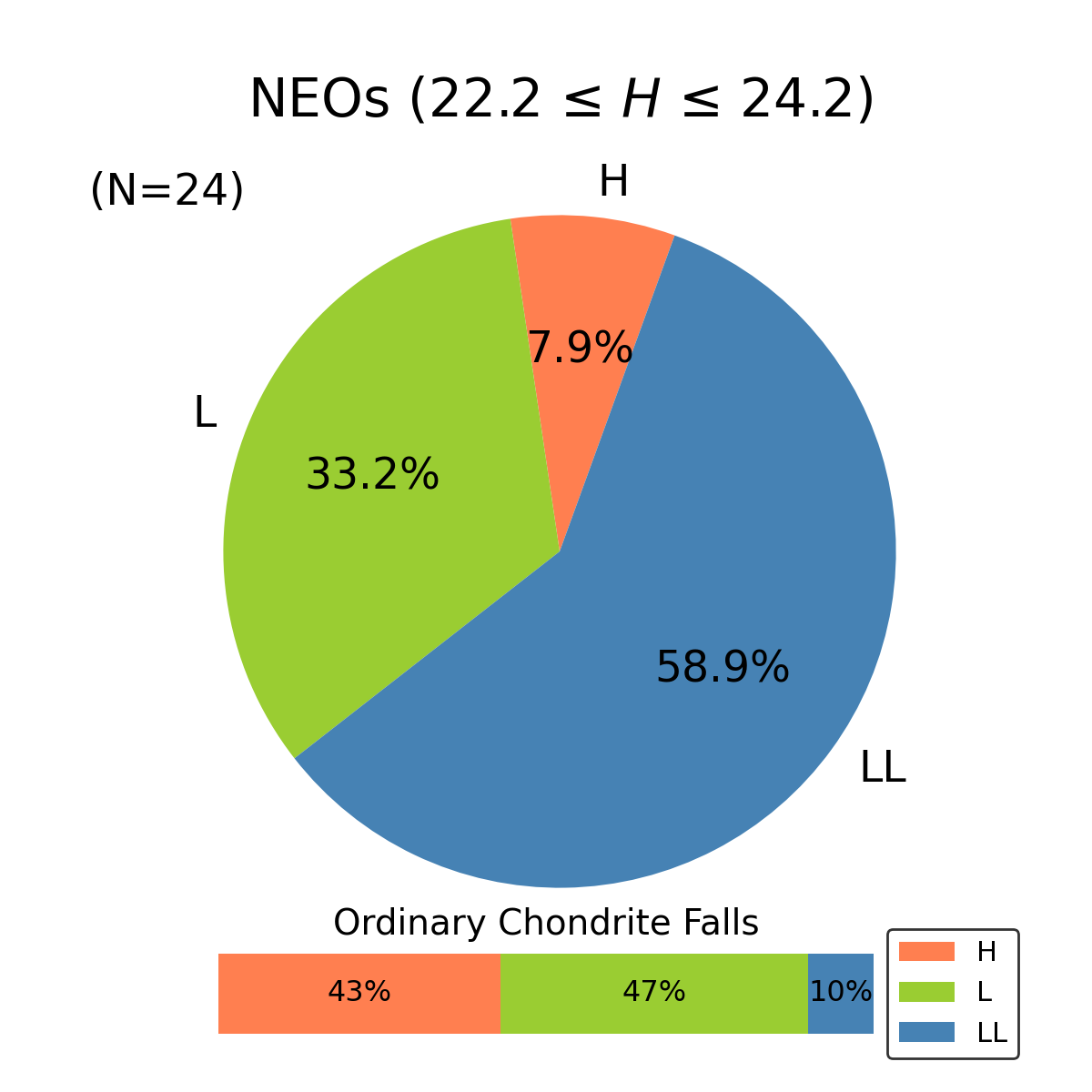}
\\
\includegraphics[width=7cm,angle=0]{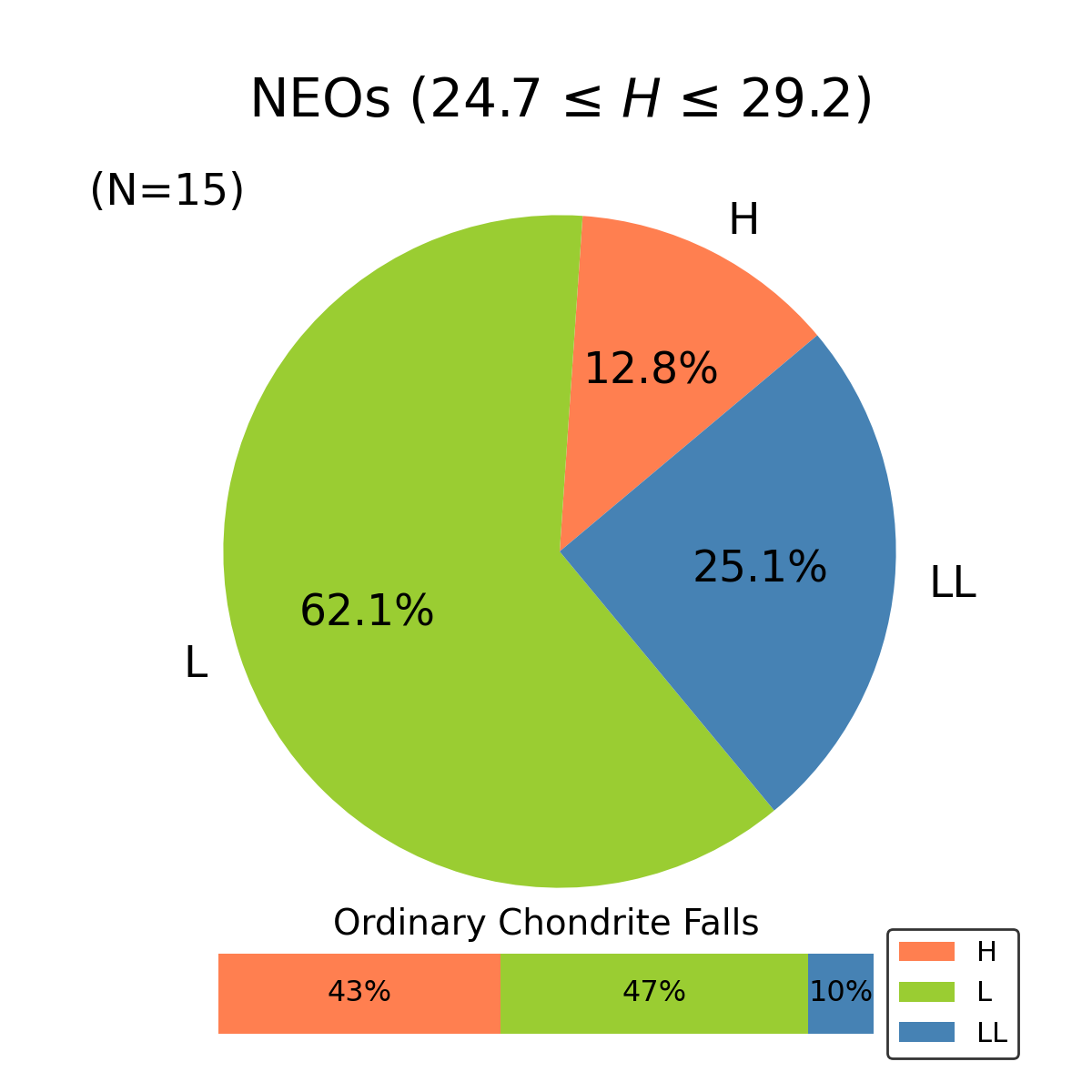}
\caption{Mean values for the ordinary chondrite subtypes for the three-subgroup analysis. The number of objects for each subgroup is indicated. The fractions of ordinary chondrite falls are shown in the horizontal bar.}
\label{f:prob_pie_charts_three_bins}
\end{center}
\end{figure}

 \begin{figure} 
\begin{center}
\includegraphics[width=8.5cm,angle=0]{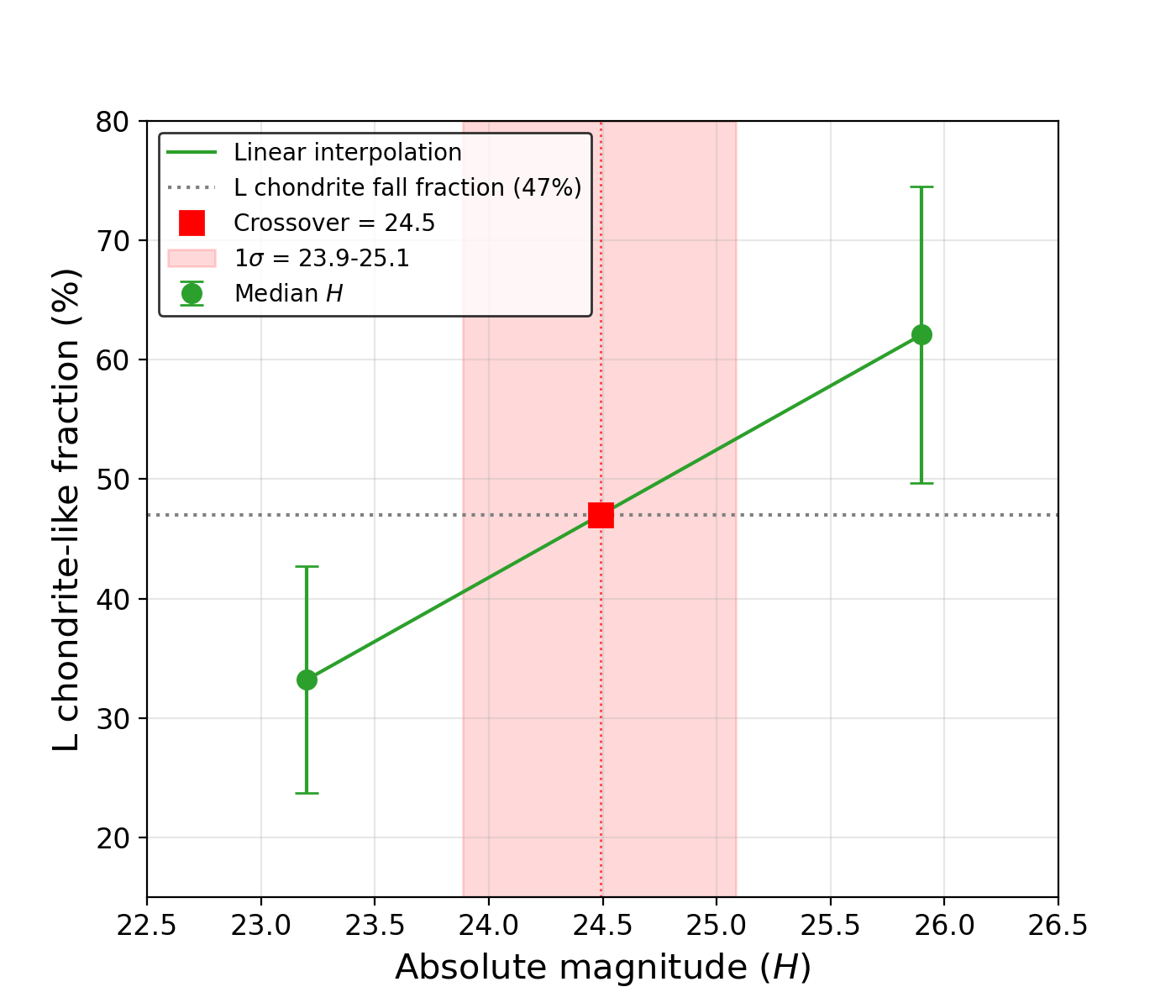}
\includegraphics[width=8.5cm,angle=0]{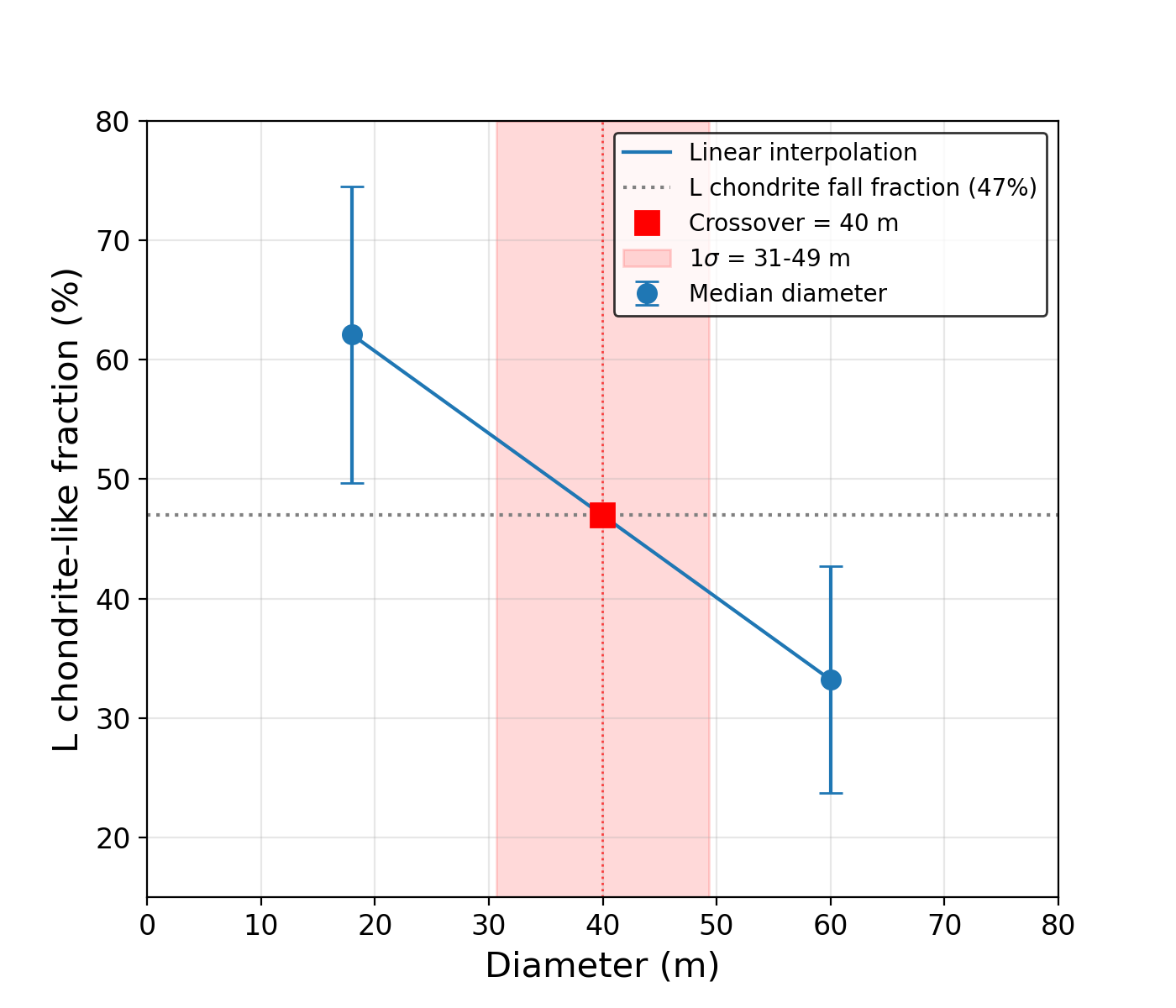}
\caption{L chondrite-like fraction vs. absolute magnitude (left) and diameter (right). A linear interpolation was performed between the median values of the 22.2 $\leq$ $H$ $\leq$ 24.2 and $24.7 \leq H \leq 29.2$ subgroups.}
\label{f:linear_interp}
\end{center}
\end{figure}

Conversely, the 22.9 $\leq$ $H$ $\leq$ 29.2 subgroup 
shows a fraction of L chondrite-like NEOs that matches remarkably well with the proportion of L chondrite meteorite falls (47\%). This close match tells us that 
the size range of $\sim$4-70 m is precisely where the pre-atmospheric parent bodies of L chondrite meteorites reside. Furthermore, it validates the core 
hypothesis of this study, i.e., that smaller NEOs have compositions more consistent with meteorite falls. 

Although fewer, there 
are still too many asteroids with LL chondrite-like compositions in the 22.9 $\leq$ $H$ $\leq$ 29.2 subgroup compared to 
LL chondrite meteorite falls (10\%). These results imply that the size-composition convergence toward meteorite-fall 
proportions requires objects with sizes smaller than $\sim$39 m, the median size for this subgroup.

Having identified a size dependence in the composition of the NEOs, we use the three-subgroup analysis to further constrain the sizes of the 
parent bodies (Table 2). As shown in Figure \ref{f:prob_pie_charts_three_bins}, there is little change in the proportion of H chondrite-like objects between the 20.0 $\leq$ $H$ $\leq$ 22.1 (7.4$\pm$4.2\%) and the 22.2 $\leq$ $H$ $\leq$ 24.2 (7.9$\pm$5.7\%)
subgroups, and only a small increase to 12.8$\pm$8.8\% in the $24.7 \leq H \leq 29.2$ subgroup. This is still substantially below the proportion of H chondrite falls, implying that the 
pre-atmospheric precursors of H chondrite meteorites have not yet been reached in size. In other words, the objects that actually enter Earth's atmosphere and produce 
H chondrite falls are predominantly smaller than $\sim$18 m (the median size for the $24.7 \leq H \leq 29.2$ subgroup) and are being sourced from a population of very 
small NEOs that remains almost entirely unsampled by current spectroscopic surveys. 

There is only a small increase in the fraction of L chondrite-like NEOs from the 20.0 $\leq$ $H$ $\leq$ 22.1 subgroup (27.2$\pm$7.1\%) to the 22.2 $\leq$ $H$ $\leq$ 24.2 subgroup (33.2$\pm$9.5\%), but then the proportion of these objects rises to 62.1$\pm$12.4\% in the $24.7 \leq H \leq 29.2$ subgroup. This means that the 
proportion of L chondrites in the $24.7 \leq H \leq 29.2$ subgroup lies above the 47\% fall fraction, while the 22.2 $\leq$ $H$ $\leq$ 24.2 
subgroup lies below it, bracketing the matching size between the central sizes of these two bins. To estimate the crossover of the absolute magnitude and the diameter, we performed a linear 
interpolation between the medians of these two subgroups. The 1$\sigma$ uncertainties in the L chondrite fractions were propagated using a Monte Carlo 
simulation to obtain a confidence interval for the crossovers (Figure \ref{f:linear_interp}). The linear interpolation places the crossover at 
$H = 24.5$ and a size of 40 m, with a 1$\sigma$ confidence range of $\sim$23.9-25.1 for $H$ and $\sim$31-49 m for the diameter. These results imply that the 
pre-atmospheric parent bodies of L chondrite meteorites are predominantly in the $\sim$18-60 m range, with a typical size of $\sim$31-49 m. The matching proportions of L chondrite-like NEOs and meteorite falls also indicate that S-complex NEOs smaller than $\sim$40 m ($H$ $>$ 24.5) are dominated by L 
chondrite-like compositions. For larger NEOs (extending to kilometer-sized bodies), LL chondrite-like compositions are dominant.

There are two caveats in this analysis that are worth mentioning: first, the analysis assumes that the L chondrite fraction varies monotonically with size between 
the two bins, an assumption supported by the trend observed in the full sample, but not strictly proven. Second, "parent body size" here means the size at which the NEO compositional mix 
matches the meteorite fall statistic, i.e., it is a central tendency at the population level, not a hard upper or lower cutoff. Real meteorite precursors will be drawn from a size 
distribution that extends somewhat beyond this range, and we are identifying where the bulk of them sit.

The persistence of an overabundance of LL chondrite-like objects in the $24.7 \leq H \leq 29.2$ subgroup (25.1$\pm$11.3\%), despite the clear trend toward L chondrite 
proportions, indicates that the size threshold below which these NEOs approach their true meteoritic proportion (10\%) has still not been 
reached, even at these small sizes. This suggests that the size-dependent depletion of LL chondrite-like bodies is a gradual, progressive trend rather than a 
sharp transition, and it has not run its full course by $\sim$4-31 m. As with the parent bodies of H chondrites, we set an upper size limit of $\sim$18 m for these objects. From this upper limit it follows that objects like the pre-atmospheric Chelyabinsk, with an estimated diameter of $\sim$20 m, are probably rare among LL chondrites.

\section{Source Regions of Small NEOs} \label{sec:source}

\begin{figure}[!h]
\begin{center}
\includegraphics[height=9.5cm]{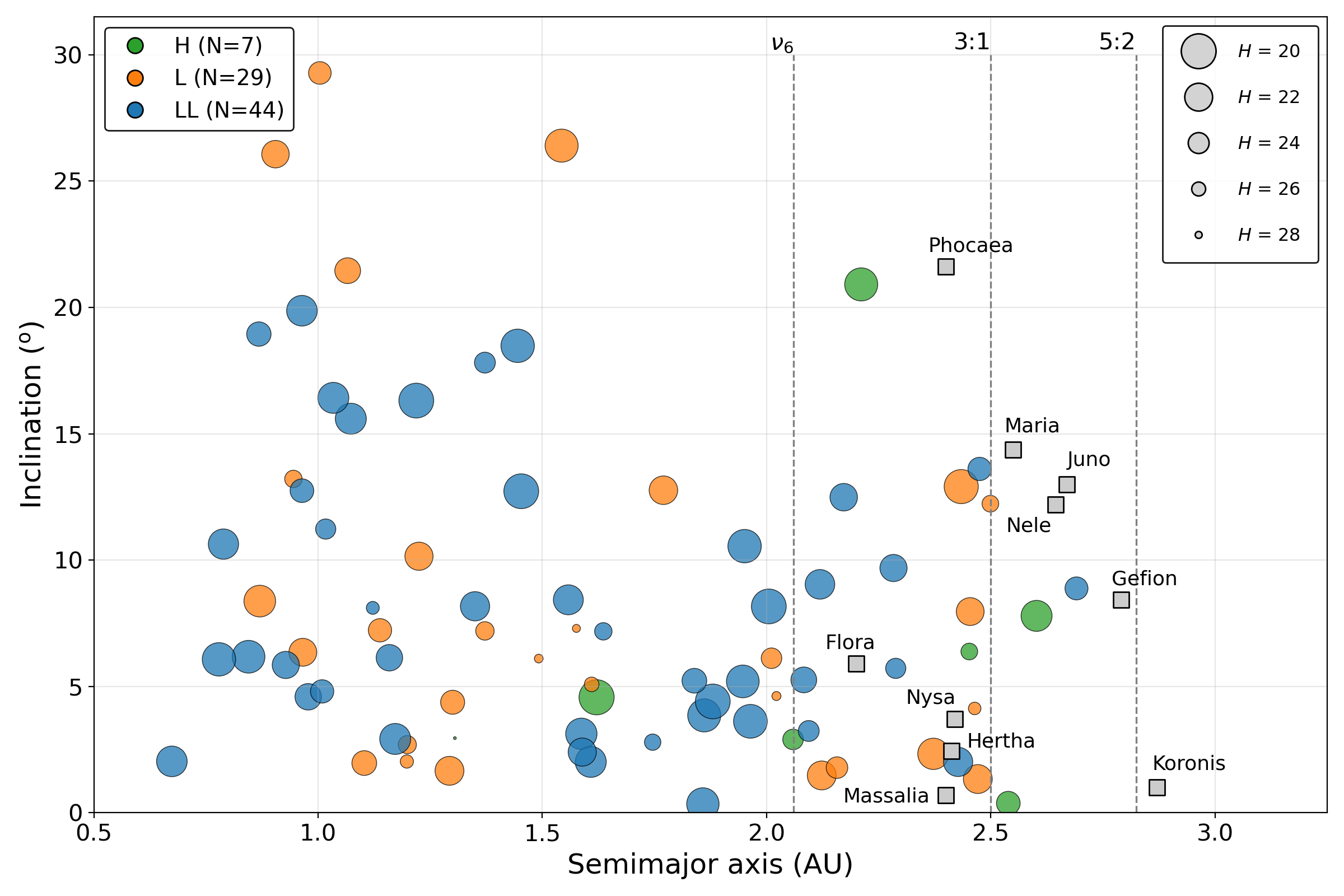}
\caption{\label{f:i_a_NEOs}{\small Inclination vs. semimajor axis for the NEOs. The main asteroid families that have been linked to 
ordinary chondrites are represented by gray squares. The $\nu_{6}$ secular resonance with Saturn and the 3:1 and 5:2 mean motion resonances with Jupiter 
are represented with dashed lines.}}
\end{center}
\end{figure}

The orbital distribution for the NEOs is shown in Figure \ref{f:i_a_NEOs}. The objects in our sample have inclinations 
$0.35^{o}\leq i \leq 29.28^{o}$ and semimajor axes $0.67\leq a \leq 2.69$ au, with most of them having $i<15^{o}$ and $a<$ 2.5 au. This Figure also shows the location of some of the asteroid families that have been associated with ordinary chondrites. An apparent concentration of the smallest L 
chondrite-like NEOs is observed at semimajor axes below the $\nu_{6}$ resonance, consistent with the prediction of \cite{2024Natur.634..561M}. However, this trend should be interpreted with caution due to the limited sample size. 

As we have already seen, establishing genetic links between NEOs and asteroid families is an active area of research that sometimes produces conflicting results 
\citep[e.g.,][]{2025M&PS...60..928J, 2026AJ....171..169V}. For this reason, in this work, instead of trying to link the NEOs to specific families, we have adopted a more general approach 
where we investigate their possible escape regions from the asteroid belt into the near-Earth space.

\cite{2019Icar..324...41B} reported systematic differences in the most common source regions for NEOs spectrally classified as H, L, and LL chondrites. Although the primary sources for these objects were all the same ($\nu_{6}$ and 3:1 resonances and the Hungaria region), the authors found that H chondrite-like NEOs show higher contributions from the 3:1 and 5:2 resonances and the Phocaea region compared to the full sample. L chondrite-like objects showed no significant deviation from the overall trend, while LL chondrite-like NEOs exhibited an enhanced contribution from the $\nu_{6}$ resonance.

The sample studied by \cite{2019Icar..324...41B} had a median diameter of $\sim$700 m, with most NEOs having sizes larger than 100 m. As discussed previously, this size is too large to be representative of 
the immediate precursors of meteorites that fall on Earth. Therefore, in this section we investigate whether sub-100 m objects follow different delivery routes from those of larger 
NEOs. The sample used includes all objects in the 22.9 $\leq$ $H$ $\leq$ 29.2 subgroup plus seven objects from 
the 20.0 $\leq$ $H$ $\leq$ 22.6 subgroup with sizes smaller than 100 m (40 NEOs in total). Probability distribution functions for the NEOs were generated using the latest 
version of NEOMOD \citep{2023AJ....166...55N, 2024Icar..41115922N}, a 12 region model that includes eight resonances ($\nu_{6}$, 3:1, 5:2, 7:3, 8:3, 9:4, 
11:5, and 2:1), an inner belt source, two high-inclination sources (Hungarias and Phocaeas), and Jupiter family comets (JFC). The model, which is implemented in the NEOMOD3 
simulator\footnote{https://www2.boulder.swri.edu/\textasciitilde{}davidn/NEOMOD\_Simulator/}, yields the probability that an object escaped out of one of these 
12 regions.

\begin{figure}[!h]
\begin{center}
\includegraphics[height=9cm]{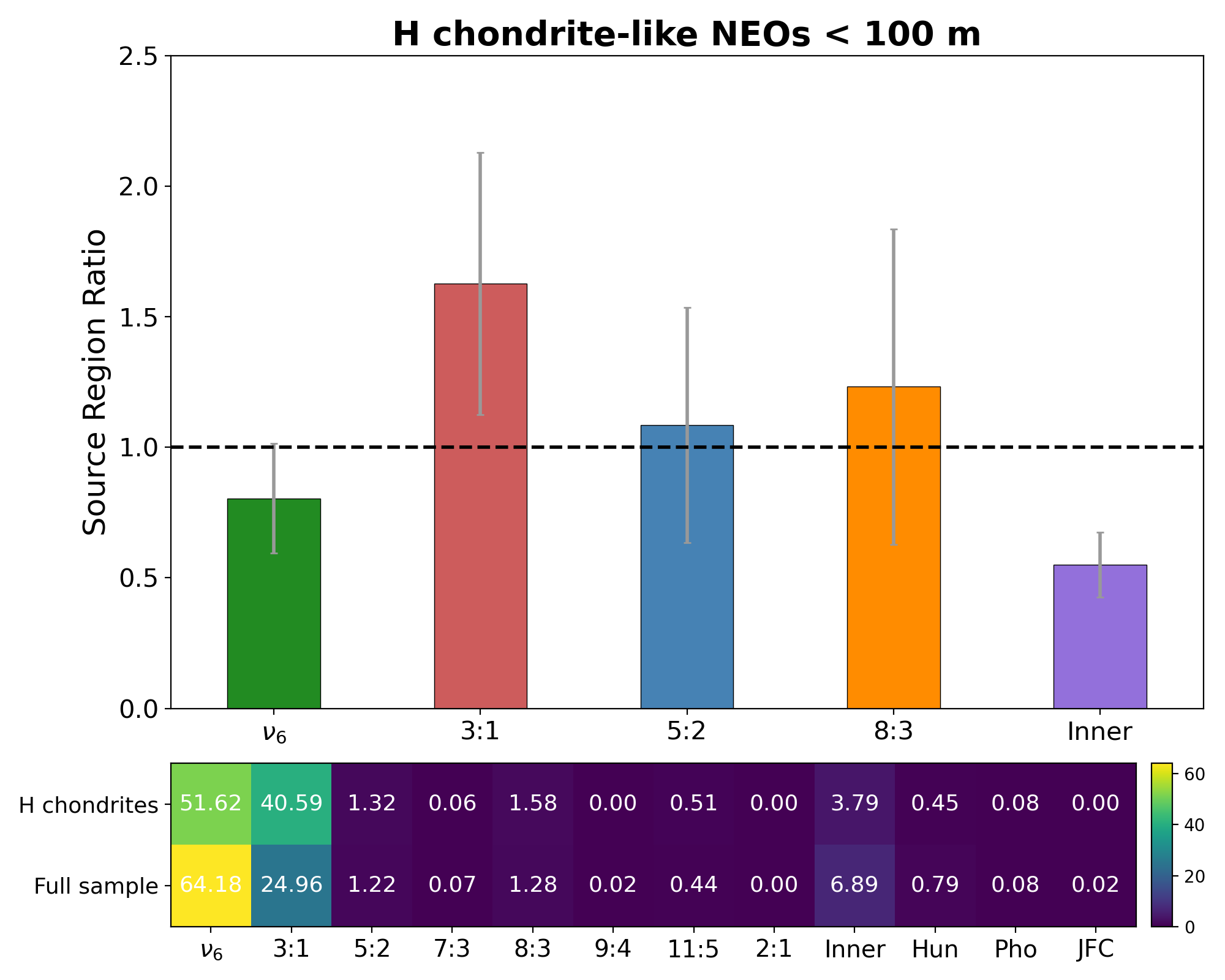}
\caption{\label{f:H_source}{\small Top: source region ratios of H chondrites relative to the full sample of NEOs. For clarity, only source regions with 
probabilities greater than 1\% in the full sample are shown. Bottom: heatmap showing the mean probability distribution 
functions of source regions for the full sample (lower row) and NEOs with H chondrite-like compositions (upper row).}}
\end{center}
\end{figure}

\begin{figure}
\begin{center}
\includegraphics[height=9cm]{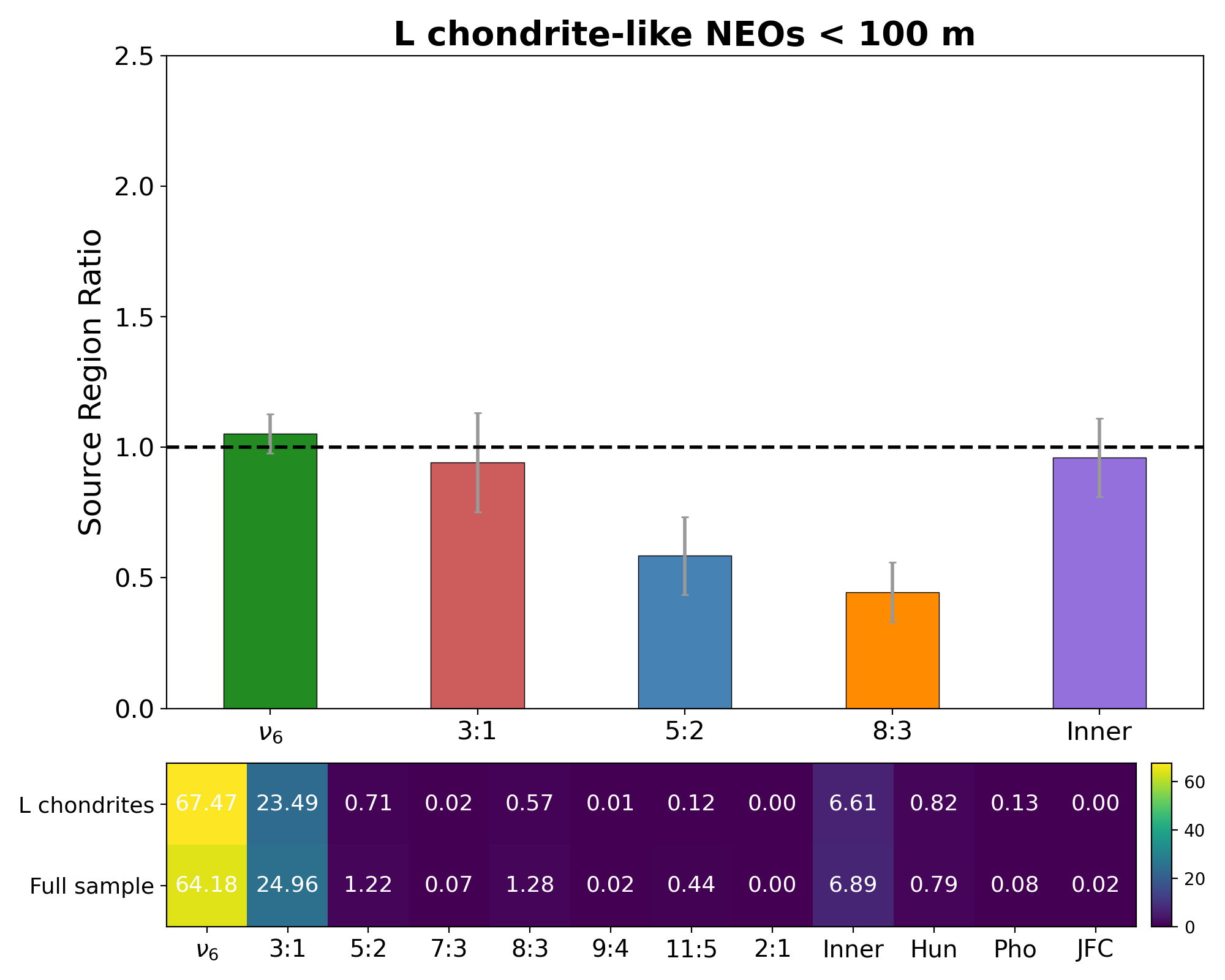}
\caption{\label{f:L_source}{\small Top: source region ratios of L chondrites relative to the full sample of NEOs. For clarity, only source regions with 
probabilities greater than 1\% in the full sample are shown. Bottom: heatmap showing the mean probability distribution 
functions of source regions for the full sample (lower row) and NEOs with L chondrite-like compositions (upper row).}}
\end{center}
\end{figure}

\begin{figure}
\begin{center}
\includegraphics[height=9cm]{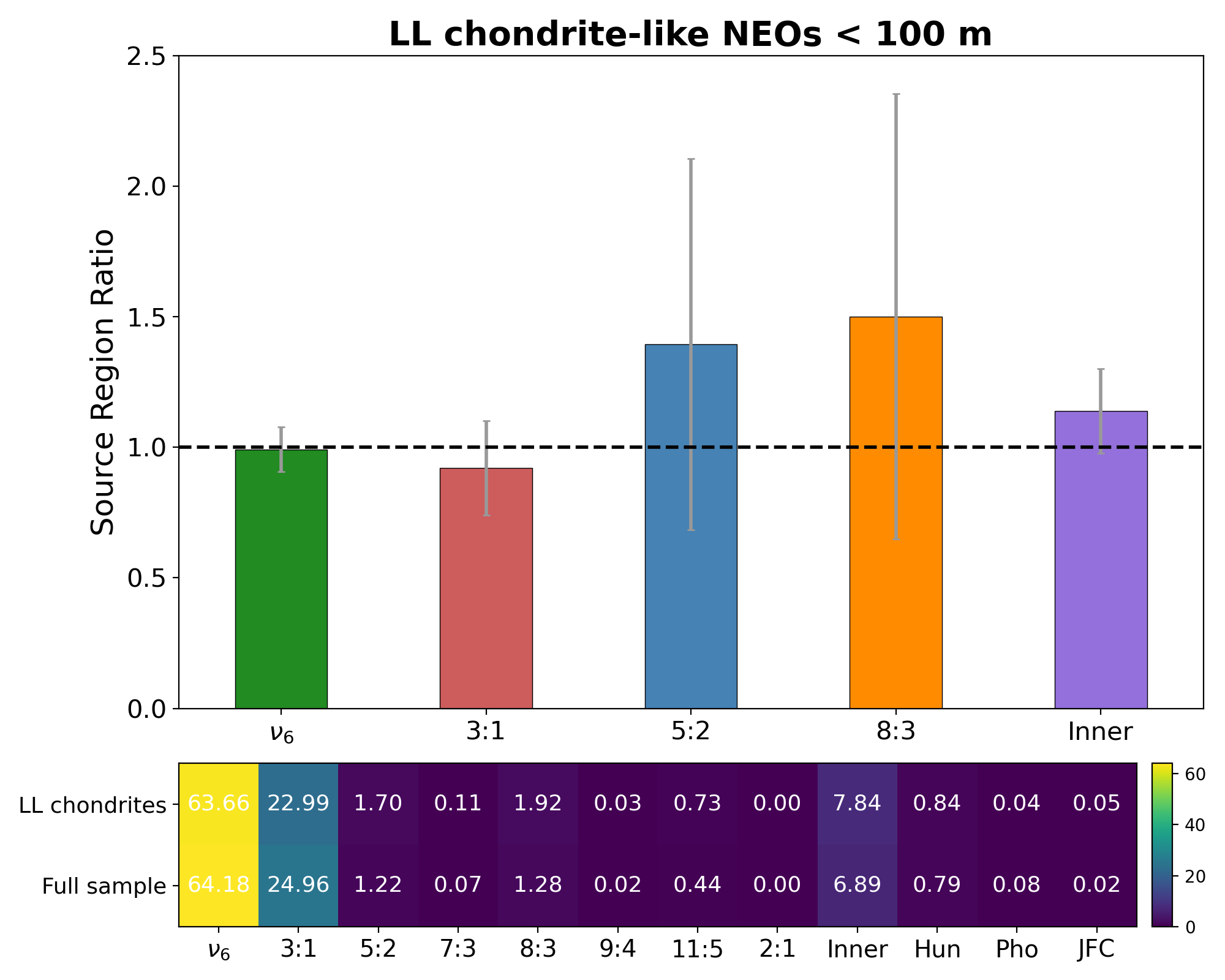}
\caption{\label{f:LL_source}{\small Top: source region ratios of LL chondrites relative to the full sample of NEOs. For clarity, only source regions with 
probabilities greater than 1\% in the full sample are shown. Bottom: heatmap showing the mean probability 
distribution functions of source regions for the full sample (lower row) and NEOs with LL chondrite-like compositions (upper row).}}
\end{center}
\end{figure}

The mean probability distribution functions of the source regions are represented as heatmaps in Figures \ref{f:H_source}-\ref{f:LL_source}. For the full 
sample of 40 NEOs (lower rows), the highest contribution comes from the $\nu_{6}$ resonance ($\sim$64\%), followed by the 
3:1 resonance ($\sim$25\%) and the inner belt source ($\sim$7\%). \cite{2019Icar..324...41B} reported lower values for the $\nu_{6}$ ($\sim$60\%) and 
3:1 (15\%) resonances, and a 15\% contribution from the Hungaria region, which is higher than the mean value obtained for our sample (0.8\%). These results 
are expected due to a size dependence in the contribution of the source regions to the NEO population \citep{2023AJ....166...55N}. In particular, the 
$\nu_{6}$, 3:1 and 8:3 resonances contribute more to smaller NEOs, whereas the other source regions have the opposite behavior \citep{2023AJ....166...55N}. It is also worth noting that the use of different models could also be partly responsible for the different results, since \cite{2019Icar..324...41B} used the seven-region model 
of \cite{2017A&A...598A..52G}. 

To determine the source region probability distribution function for the ordinary chondrite subtypes, we followed the same procedure described in 
\cite{2019Icar..324...41B}. For each subtype, weighted means were calculated by convolving the probability distribution functions obtained from the 
machine learning classifier and NEOMOD. The weighted mean probability distribution functions for H, L, and 
LL chondrite-like NEOs are shown in the upper rows of the heatmaps (Figures \ref{f:H_source}-\ref{f:LL_source}). Next, following \cite{2019Icar..324...41B}, 
we calculated the ratios between the mean probability distribution functions of each subtype and the full sample (upper rows/lower rows). This 
procedure allows us to measure the contribution of the source regions relative to a baseline, where a source region ratio equal to one (dashed line) indicates that there 
are no changes relative to the full sample. Error bars were generated calculating the standard error of the mean. 
Source region ratios for H, L, and LL chondrites are shown in the top panel of Figures \ref{f:H_source}, \ref{f:L_source}, and  
\ref{f:LL_source}, respectively. For clarity, only source regions with probabilities greater than 1\% in the full sample are shown.

In general, the three ordinary chondrite subtypes do not deviate substantially from the results obtained for the full sample, where the major contributions come from the 
$\nu_{6}$ and 3:1 resonances and the inner belt source. However, some variations in the source region ratios were observed. For NEOs with H chondrite-like compositions (Figure \ref{f:H_source}), the 3:1 resonance shows a higher contribution than the full sample, consistent with the results found for NEOs larger than 100 m \citep{2019Icar..324...41B}; these objects also show a lower contribution from the inner belt source relative to the baseline. For L chondrite-like NEOs 
(Figure \ref{f:L_source}), a drop in the contribution of the 5:2 and 8:3 resonances compared to the full sample was found. LL chondrite-like objects show an increased contribution from the 5:2 and 8:3 resonances and the inner belt source (Figure \ref{f:LL_source}); however, this increase falls within the error bars.

\section{Summary} \label{sec:Summ}

This study examined the relationships among size, composition, and source region for S-complex NEOs. Our findings largely resolve the long-standing discrepancy 
between the overabundance of LL chondrite-like NEOs and the paucity of LL chondrites among ordinary chondrite falls. Our most significant results are:

\begin{itemize}

\item We found that for NEOs in the $24.7 \leq H \leq 29.2$ subgroup, the proportion of H chondrite-like objects (12.8$\pm$8.8\%) is too low compared 
to H chondrite meteorite falls (43\%), suggesting that the majority of the pre-atmospheric precursors of these meteorites have sizes smaller than $\sim$18 m (the median size).

\item The fraction of L chondrite-like NEOs with absolute magnitudes of $23.9\leq H \leq 25.1$ matches the proportion of L chondrite falls (47\%), implying 
that the size range of $\sim$31-49 m is where most of the pre-atmospheric parent bodies of these meteorites reside.

\item The fraction of asteroids with LL chondrite-like compositions in the $24.7 \leq H \leq 29.2$ subgroup (25.1$\pm$11.3\%) exceeds the proportion of LL chondrite falls (10\%), indicating that the size threshold below which these NEOs approach their meteoritic proportion has still 
not been reached, and must lie below $\sim$18 m.

\item For sub-100 m S-complex NEOs, we found that the main source region is the $\nu_{6}$ resonance, followed by the 3:1 resonance and the inner belt source.

\end{itemize}

\begin{acknowledgments}

This research work was supported by NASA Yearly Opportunities for Research in Planetary Defense grant 80NSSC22K0514 (PI: V. Reddy). We thank the IRTF TAC for awarding time to this project, and to the IRTF TOs and MKSS staff for their support. The authors wish to 
recognize and acknowledge the very significant cultural role and reverence that the summit of Maunakea has always had within the indigenous Hawaiian community. We are 
most fortunate to have the opportunity to conduct observations from this mountain. Taxonomic type results presented in this work were determined, in whole or in part, using a Bus-DeMeo Taxonomy Classification Web tool by Stephen M. Slivan, developed at MIT with the support of National Science Foundation grant 0506716 and NASA grant 
NAG5-12355. We thank the anonymous reviewers for their helpful comments, which helped improve this paper.

\end{acknowledgments}

\clearpage

\begin{appendix}

\vspace{-6mm}

\begin{figure*}[!ht]
\begin{center}
\includegraphics[width=16cm,angle=0]{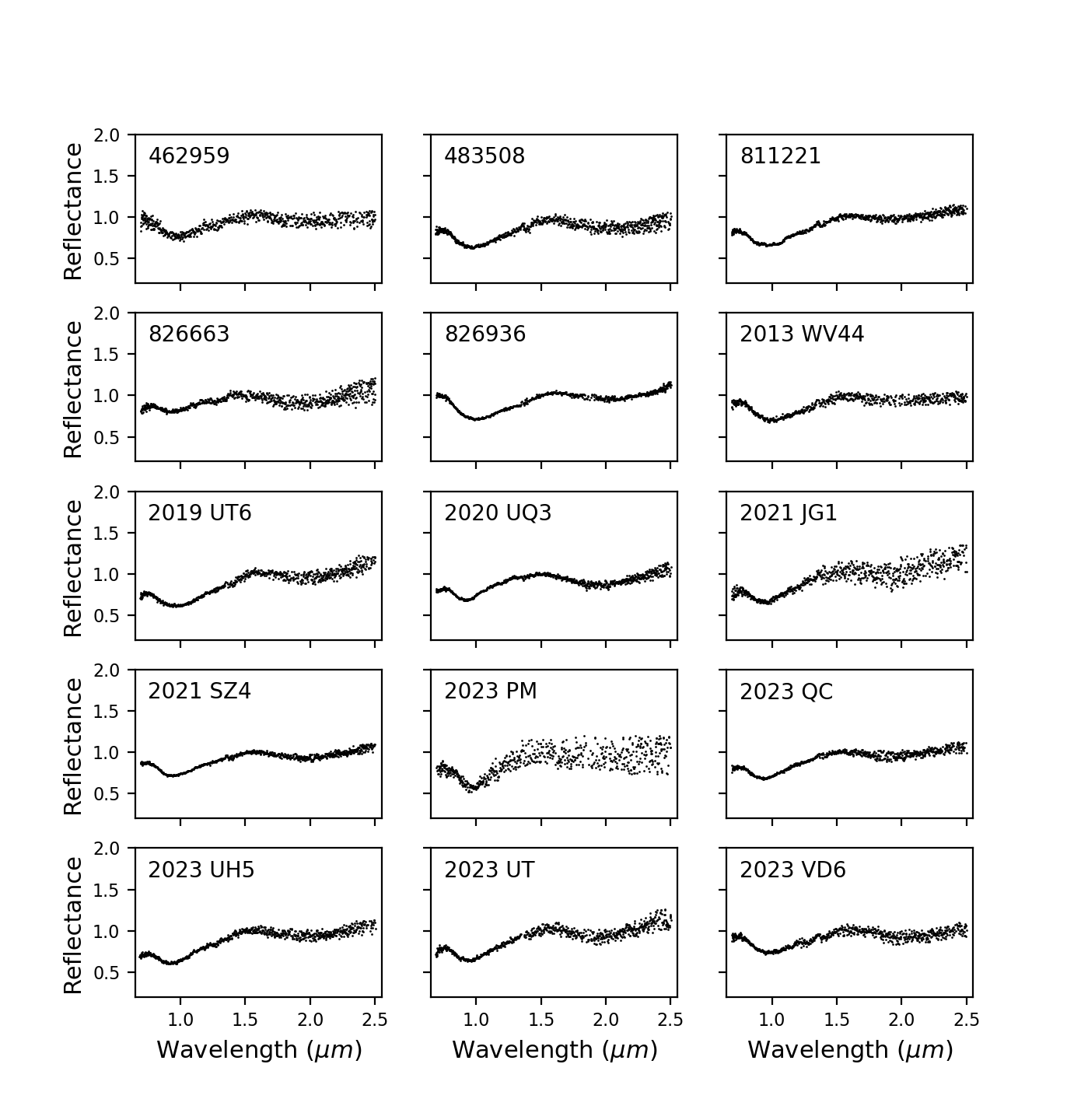}
\end{center}
\caption{Near-IR spectra of the new 25 NEOs included in this study.}
\label{f:all_spectra_1}
\end{figure*}

\begin{figure*}[!ht]
\begin{center}
 \includegraphics[width=16cm,angle=0]{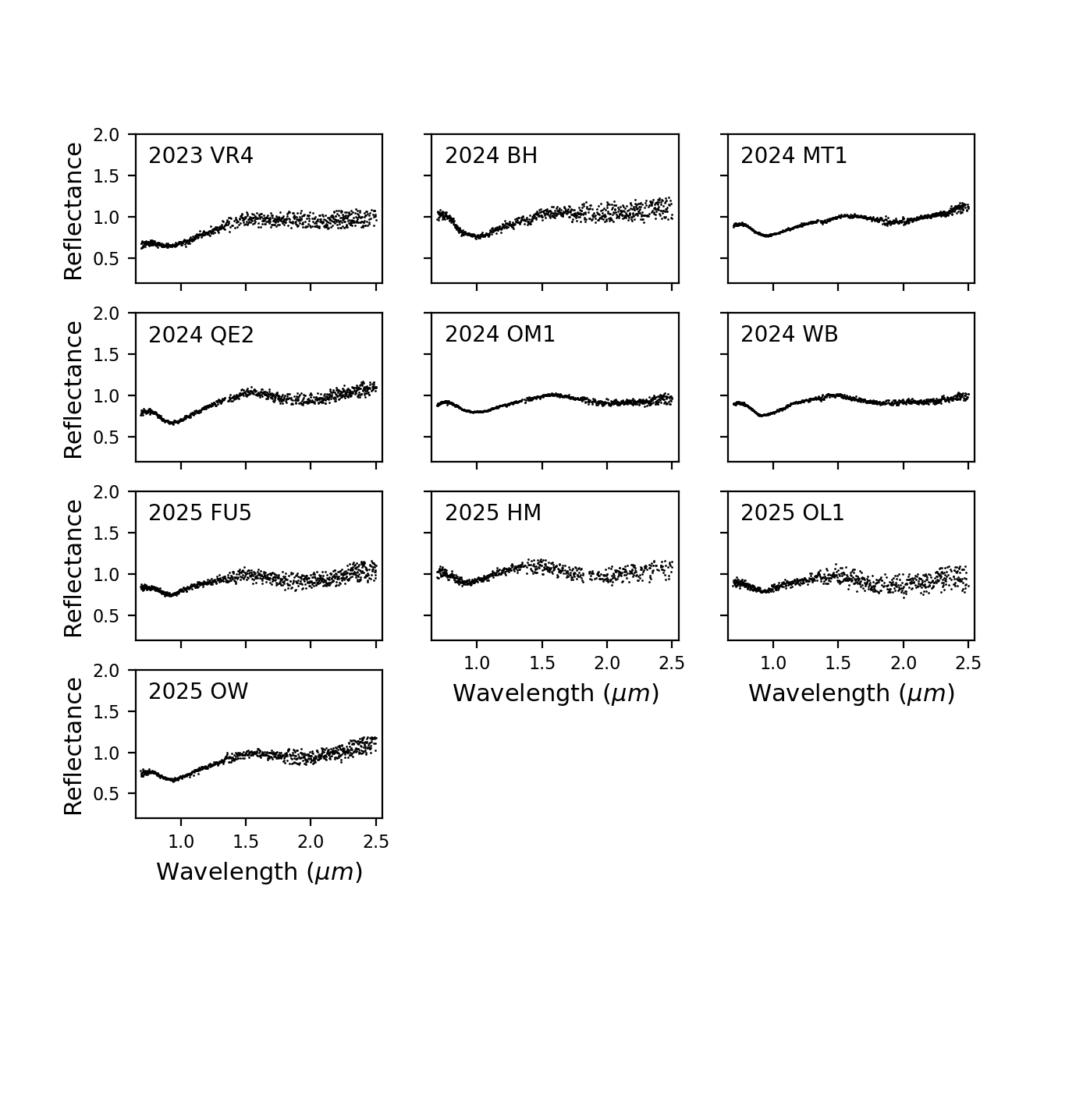}
 \end{center}
 \caption{Near-IR spectra of the new 25 NEOs included in this study.}
\label{f:all_spectra_2}
\end{figure*}

NIR spectra obtained with the SpeX instrument on the IRTF between May-2021 and July-2025 are reported in Figures \ref{f:all_spectra_1} and \ref{f:all_spectra_2}.

\vspace{-3mm}

\end{appendix}

\end{document}